\documentclass[aps,prfluids,reprint,superscriptaddress,floatfix,onecolumn]{revtex4-2}
\usepackage{amssymb}
\usepackage{soul}
\usepackage{multirow}
\usepackage[german, english]{babel}
\usepackage[T1]{fontenc}
\usepackage{lmodern}
\usepackage{xcolor}
\usepackage{graphics}
\usepackage{rotating}   
\usepackage{verbatim} 
\usepackage[ansinew]{inputenc}
\usepackage{orcidlink}
\usepackage{graphics}
\usepackage{graphicx,mathtools}
\usepackage{anyfontsize}
\usepackage{nicefrac}
\usepackage{lmodern} 
  
\renewcommand{\small}{\fontsize{10}{12}\selectfont}   
\usepackage[a4paper,top=2cm,bottom=2cm,left=2.3cm,right=2.3cm]{geometry}
\usepackage{setspace}
\usepackage{listings,xcolor}
\definecolor{darkblue}{rgb}{0,0,.5}
\hypersetup{colorlinks=true, breaklinks=true, citecolor=blue, linkcolor=blue, menucolor=darkblue, urlcolor=blue}

\newcommand{\fett}[1]{\boldsymbol{#1}}
\newcommand{\dd}{{\rm{d}}}
\newcommand{\ii}{{\rm{i}}}
\newcommand{\be}{\begin{equation}}
\newcommand{\ee}{\end{equation}}
\newcommand{\mPsi}{{\mathit \Psi}}
 
\newcommand{\smallL}{\mathchoice{\scriptstyle \mathrm{L}}{\scriptstyle \mathrm{L}}{\scriptscriptstyle \mathrm{L}}{\scriptscriptstyle \mathrm{L}}}
\newcommand{\nuL}{\nu_{\mskip1mu\smallL}}
\newcommand{\MR}{\mathchoice{\scriptstyle \mathrm{MR}}{\scriptstyle \mathrm{MR}}{\scriptscriptstyle \mathrm{MR}}{\scriptscriptstyle \mathrm{MR}}}
\newcommand{\DSE}{\mathchoice{\scriptstyle \mathrm{DS, E}}{\scriptstyle \mathrm{DS, E}}{\scriptscriptstyle \mathrm{DS, E}}{\scriptscriptstyle \mathrm{DS, E}}}

\definecolor{darkgreen}{rgb}{0,0.5,0}

\begin{document}

\hfill {\scriptsize RBI-ThPhys-2026-01}

\title{Complex-time singular structure of the 1D Hou--Luo model}

\author{Cornelius Rampf\orcidlink{0000-0001-5947-9376}}
\email{cornelius.stefan.rampf@irb.hr} 
\affiliation{Division of Theoretical Physics, Ru\dj er Bo\v{s}kovi\'c Institute, Bijeni\v{c}ka cesta 54, 10000 Zagreb, Croatia}

\author{Sai Swetha Venkata Kolluru\orcidlink{0009-0009-4672-7874}}
\email{venkata-saiswetha.kolluru@cea.fr}
\affiliation{SPEC, CEA Saclay, Gif-sur-Yvette, F-91191, France}
\affiliation{Centre for Condensed Matter Theory, Department of Physics, Indian Institute of Science, Bengaluru, India - 560012}

\date{\today}

\begin{abstract}
Starting from smooth initial data, we investigate the complex-time analytic structure of the one-dimensional Hou--Luo (HL) model, a wall approximation of the three-dimensional axisymmetric Euler equations. While the finite-time blow-up in this setting has been already established, here we chart the entire singular landscape. This analysis is enabled by a novel formulation of the HL model in Lagrangian coordinates, in which the time-Taylor coefficients of the flow fields are evaluated symbolically to high truncation order. Our results are threefold. First, we show that the Lagrangian series for the vorticity converges within the complex-time disc of radius~$t_\star >0$ and is free from (early-time) resonances that impede the Eulerian formulation. Second, applying asymptotic analysis on the series, we recover both the blow-up time and the singularity exponent with high accuracy. This also enables a quantitative assessment of the Beale--Kato--Majda criterion, which we find correctly identifies the blow-up time, but washes out the local singularity exponent, as it relies on a spatial supremum. Third, and most importantly, we develop a Lagrangian singularity theory that predicts the eye-shaped singularity profile observed in Eulerian coordinates by exploiting the driving mechanism of the blow-up: The accumulation of multiple fluid particles at the same Eulerian position. The employed techniques extend recently introduced methods for the inviscid Burgers equation [C.\ Rampf {\it et al.}, \href{https://doi.org/10.1103/PhysRevFluids.7.104610}{Phys.\ Rev.\ Fluids {\bf 7} (2022) 10, 104610}], and can be further adapted to higher spatial dimensions or other hydrodynamical equations.

\end{abstract}

\maketitle

\section{Introduction}

The proof of global regularity for smooth solutions of hydrodynamic partial differential equations is a core problem in mathematical fluid dynamics. For the three-dimensional (3D) Navier--Stokes and Euler equations in particular, the formation of finite-time singularities signals the breakdown of global regularity and may have plausible implications for understanding anomalous dissipation in Navier--Stokes turbulence~\cite{Eyink-onsager}. In this context, reduced or symmetry-constrained settings---such as axisymmetric flows in bounded domains---have emerged as valuable testing grounds for exploring singular behavior. Notably, the 3D axisymmetric incompressible Euler (3DAE) equations in a wall-bounded domain have attracted renewed interest in recent years, following numerical evidence for a finite-time blow-up~\cite{HouLuo2014}.

To detect singular behavior in hydrodynamical equations, a variety of analytical and numerical criteria have been devised. A prominent example is the Beale--Kato--Majda (BKM) criterion~\cite{BKM-1984}, which links the blow-up of 3DAE to the divergence of the time integral of its maximum vorticity~$\| \fett \omega \|_\infty$. Non-blow-up criteria have been developed as well, such as Ref.\,\cite{Deng-Hou-Yu-2004math.ph...2032D} who identified geometric and dynamical constraints on vorticity growth that preclude singularity formation. Related approaches---often based on controlling quantities tied to vortex stretching,  alignment, pressure or energy---have been explored in closely related models, providing valuable insights into the mechanisms that may lead to or prevent a blow-up~\cite{HouRuo2006JNS....16..639H,Barkley2020,Fehn2022JFM...932A..40F}.

Complementary to such norm-based criteria, analyticity-based methods probe singular behavior through the loss of spatial or temporal analyticity of the solutions. In the spatial case, the analyticity-strip method~\cite{sulem} tracks the distance of the nearest complex singularities from the real spatial domain by monitoring the exponentially decaying tail of (Fourier) spectral coefficients \cite{brachet,Brachet1992PhFlA...4.2845B,cichowlas}. For the 3DAE (and related problems), Ref.\,\cite{Kolluru2022PhRvE} used Fourier--Chebyshev pseudospectral methods and the analyticity-strip method to trace the movement of the convergence-limiting singularities in complex space to estimate the time of singularity. However, a precise determination of the blow-up time was obscured by the appearance of oscillatory artifacts called {\it tygers}, a phenomenon also observed in other Eulerian fluid simulations~\cite{Ray2011,VenkataramanRay2017RSPSA.47360585V,Leoni2018PhRvF...3a4603C,2014Banerjee,2023Murugan} (but see \cite{KolluruBessePandit2024}).

Another approach to detect temporal singularities is to consider temporal analyticity \cite{Taylor-Green-1937RSPSA.158..499T}. There, one constructs time-Taylor series for the flow fields and searches for complex-time singularities either via asymptotic analysis or by using convergence-accelerated methods such as Pad\'e approximants (see e.g.\ \cite{vanDyke1974}). Combined with high-precision arithmetic, these methods have been used to study the analytic structure of Euler flows~\cite{Dyke-1975SJAM...28..720V,Morf1980,1983Brachet,1984Brachet,1997Pelz}, turbulent Navier--Stokes flows~\cite{1981Morf}, and related hydrodynamical problems~\cite{frisch-morf,1982Meiron,2000Ohkitani}. Recently, Ref.\,\cite{Rampf:2022apj} investigated time-Taylor series solutions for 1D inviscid Burgers, relying on the symbolic computation of coefficients, which avoids numerical artifacts that can arise when sampling fields on collocation points. Moreover, extrapolation procedures often require the evaluation of (ratios of) coefficients at quasi-singular points, which can be handled more robustly when exact symbolic expressions for the coefficients are available.

In this paper, we consider the 1D Hou--Luo (HL) model, originally introduced in Ref.\,\cite{HouLuo2014} and subsequently analyzed in Refs.\,\cite{Choi2017,Huang2025,2018Do} as a reduced model to capture the singular dynamics of the full 3DAE system on the wall. Specifically, we investigate the temporal analyticity of the HL system by employing various time-Taylor series solutions. Similar methods have recently been applied to the HL model by Ref.\,\cite{KolluruPandit2024}, but exclusively in Eulerian coordinates by sampling the flow fields using quadruple-precision pseudospectral methods. In contrast, we examine the analytic structure of the HL system using both Eulerian and Lagrangian coordinates, furthermore relying on the symbolic computation of all series coefficients to achieve maximal precision. Regarding the asymptotic analysis in Eulerian coordinates, we confirm and slightly extend the results of Ref.\,\cite{KolluruPandit2024},  providing approximate extrapolation results near the real-time axis, thereby revealing a continuous landscape of  singularities arranged in an eye-shaped pattern.

By reformulating the HL model in Lagrangian coordinates, we find that the analytic structure of the problem simplifies considerably. Specifically, instead of the continuous eye-shaped singularity structure observed in Eulerian coordinates, the Lagrangian formulation is limited only by two real-valued singularities, corresponding to the blow-up time in the positive and negative time directions. This simplified analytic structure in Lagrangian coordinates enables us to develop a novel singularity theory for the HL system. Conceptually, the theory is analogous to the one recently developed for the 1D inviscid Burgers equation~\cite{Rampf:2022apj}. Here, however, we generalize the theory to accommodate series solutions, rather than the exact closed-form solutions available in the Burgers case. Incorporating the resulting rounding errors entails an important conceptual extension, thereby making the singularity theory versatile to be applied to more complex hydrodynamical equations.

The remaining part of this paper is organized as follows. We begin with preliminary considerations, in particular linking the HL model to its parent system, the 3DAE. In Sections\,\ref{sec:eul_analysis}--\ref{sec:LagAnalysis}, we provide series solutions and asymptotic analysis of the HL system, respectively formulated in Eulerian and Lagrangian coordinates; see Fig.\,\ref{fig:sing} for a summary of singularity results.  Equipped with the asymptotic insights, we derive a model for the Lagrangian vorticity that, to a good approximation, resolves the spatio-temporal regime near the blow-up. This is achieved via asymptotic completion (AC) of the truncated series solution (Sec.\,\ref{sec:UV}). The resulting AC model then serves as a surrogate for a simulation output for the vorticity, allowing us to test how much local singularity information can be retrieved using the BKM condition (Sec.\,\ref{sec:BKM}). Finally, we develop the singularity theory in Sec.\,\ref{sec:sing-theory} and conclude in Sec.\,\ref{sec:concl}.

\section{Preliminary considerations and initial conditions}

The 1D Hou--Luo (HL) model was proposed by Ref.~\cite{HouLuo2014} to approximate the wall dynamics of the 3D axisymmetric Euler equations, which are conjectured to blow up after a finite time. 
The 3D axisymmetric incompressible Euler equations are given as follows:
\begin{align}\label{eq:3dae}
    \partial_t \omega^1 + u^r \partial_r \omega^1 + u^z \partial_z \omega^1 &= \partial_z (u^1)^2 \,, \qquad
    \partial_t u^1 + u^r \partial_r u^1 + u^z \partial_z u^1 = 2 u^1 \partial_z \psi^1 \,,
\end{align}    
where $\fett{u}=(u^r,u^{\theta},u^z)^\top$ is the velocity field in cylindrical coordinates. We have defined $f^1 = f^\theta/r$ for $\fett f= \fett u, \fett \omega, \fett \psi$, where $\fett \nabla \times \fett{\psi} = \fett{u}$~\cite{HouLuo2014}. 
The incompressibility condition is $\fett \nabla \cdot \fett{u}=0$, and, together with the definition of the vorticity~$\fett \omega = \fett \nabla \times \fett{u}$, this implies a Biot--Savart relation between~$\fett u$ and~$\fett \omega$. Numerical studies have shown that the secondary flows in the $r-z$ plane play a crucial role in driving the finite-time singularity, which precipitates on the symmetry line on the wall~\cite{Barkley2020}. In order to construct a 1D approximation of the singular dynamics, the Authors of Ref.\,\cite{HouLuo2014} restrict the full 3D axisymmetric Euler equations to the wall ($r=1$) and map the corresponding 3D fields to their 1D counterparts as follows: $ (u^1)^2(r=1,z) \leftrightarrow u(z),\, \omega^1 (r=1,z) \leftrightarrow w(z),\, u^z (r=1,z) \leftrightarrow v(z)$. 
{By neglecting radial derivative terms at the wall, justified and derived in Ref.\,\cite{HouLuo2014}, 
the Authors proposed the following model}
\begin{align}  \label{eqs:HL}
  \boxed{ \partial_t u + v \,\partial_z u  = 0 \,, \qquad
   \partial_t w + v \,\partial_z w = \partial_z u\,, \qquad
   \partial_z v = \mathcal{H}[w]  \,
   } 
   \,,
   \qquad\quad \text{[Hou--Luo model]}
\end{align}
with an assumed periodicity $z \in [-\pi,\pi)$.
The last equation in~\eqref{eqs:HL} is a 1D analogue of the Biot--Savart law mentioned above, for which the Hilbert transform~$\mathcal{H}$ is employed (see \cite{ConstantinLaxMajda1985} for details). Starting from smooth initial data, Ref.\,\cite{Choi2017} has mathematically proven the loss of global regularity and the resulting finite-time blow-up of the Hou--Luo (HL) model~\eqref{eqs:HL}, also addressing its connection to the 2D Boussinesq system in half-space restricted to the wall. In contrast, our approach is explicit in the sense that (a) we trace the singularities directly using Taylor-series expansions and asymptotic extrapolation methods, and (b) our methods can be straightforwardly applied to other fluids systems and/or other initial data.

Since our approach is constructive, we rely on specifying explicit initial data, for which we use
\begin{align}\label{eq:ICs}
    u(z,t=0) = \sin^2 z \,, \qquad w(z,t=0) = 0 \,.
\end{align}
With these initial conditions and using Taylor-series expansions in time $t$ up to 72nd order for the fluid variables, we test three complementary extrapolation methods yielding a blow-up time of $t = t_\star \simeq 2.118$ at locations $z = z_\star = 0, \pm \pi/2, \pm\pi$ (see Fig.\,\ref{fig:sing} for a summary of results). Similar blow-up estimates for the HL model have been provided by Ref.\,\cite{KolluruBessePandit2024}, however only for a scaled (sc) version of the above initial data, specifically setting $u^{\rm sc}(t=0) = 10^4 \sin^2(12\pi z)$ and $w^{\rm sc}(t=0)=0$; in App.\,\ref{app:MR} we repeat some of the below analysis for such scaled initial data, finding $t_\star^{\rm sc} \simeq 0.003449$ instead (which agrees with the blow-up estimate of~\cite{HouLuo2014} for 3DAE within $1.6\%$).
We remark that, for clarity of presentation, the figures below often show only half of the periodic $z$-domain, since $u$ is mirror symmetric while $v,w$ are point symmetric at $z=0$, owing to the choice of initial data~\eqref{eq:ICs}.

In Sec.\,\ref{sec:eul_analysis}, we begin with an asymptotic analysis formulated in Eulerian coordinates, largely following the methods of Ref.\,\cite{KolluruPandit2024}, while slightly extending their results. Specifically, our (still approximate) extrapolation method leads to a continuous landscape of complex-time singularities, including estimates on the blow-up time (see Fig.\,\ref{fig:sing}), while Ref.\,\cite{KolluruPandit2024} reported singularities at isolated points in complex-time plane. In Sec.\,\ref{sec:LagAnalysis}, we analyze the Lagrangian formulation of the HL model, in particular finding that the Lagrangian representation of the vorticity (and other fluid variables) is only limited by two real singularities, located at $t = \pm t_\star$, i.e., the time of blow-up. This simplified analytic structure in Lagrangian coordinates allows us {to (i) develop an ``asymptotically completed'' model with superior performance compared to the Lagrangian series (Sec.\,\ref{sec:UV}); (ii) use this model to probe the limitations of global blow-up criteria (Sec.\,\ref{sec:BKM}); and (iii) formulate a singularity theory describing the full landscape of Eulerian singularities (Sec.\,\ref{sec:sing-theory}).}

\section{Hou--Lou model in Eulerian coordinates} \label{sec:eul_analysis}

In Sec.\,\ref{sec:EulerianSeries} we solve the 1D HL model~\eqref{eqs:HL} with initial conditions~\eqref{eq:ICs} using Taylor-series expansions formulated in a new time variable $\tau =t^2$. Asymptotic analysis based on the Mercer--Roberts method is presented in~Sec.\,\ref{sec:asyEuler}. See Fig.\,\ref{fig:sing} summarizing our main results,  including those obtained from the Lagrangian theory developed in Sec.\,\ref{sec:sing-theory}.

\subsection{Series representation for the Eulerian fluid variables} \label{sec:EulerianSeries}

One way to analyze the analytic structure of the HL model is to consider Taylor-series solutions for its fluid variables. Indeed, as is well known, the radius of convergence of a Taylor series is determined by the closest singularity(ies), which serve as signs of non-analyticity. To motivate the below considerations---tailored to the specific problem at hand---it is useful to first investigate the series solutions at low orders, thereby treating the expansion formally as perturbative. Indeed, using the initial data~\eqref{eq:ICs} as well as expanding the fluid variables $f = u,v,w$ about $t=0$ according to $f= f_0 + f_1 t + f_2 t^2 + \ldots$ with $f_n =f_n(z)$, it is straightforward to find from Eqs.\,\eqref{eqs:HL} the first terms in the series solutions,
\begin{subequations}\label{eqs:eul_ttc_exp}
\begin{align}
    u(z,t) &= \sin^2(z) +  t^2 \, \sin^2(z) \cos ^2(z) + t^4 \, \frac{ \sin (2 z) \sin (4 z)}{12}  +  O(t^6)  \,, \label{eq:u-series}  \\
    v(z,t) &= - t \sin (z) \cos(z)  - t^3 \frac{\sin(4z)}{12}  - t^5   \frac{3 \sin(2z) - 2 \sin(6z)}{90}+ O(t^7)  \,, \label{eq:v-series} \\
    w(z,t) &= t \, \sin^2(z) +  t^3 \, \frac{\sin (4z)}{3} + t^5 \,  \frac{2 \sin (6 z)-\sin (2 z)}{15} + O(t^7) \,; \label{eq:w-series}
\end{align}    
\end{subequations}
see Ref.\,\cite{KolluruPandit2024} for the derivation of the corresponding recursive relations.
Observe that the series for~$u$ lacks odd time-Taylor coefficients, while~$v,w$ possess no even coefficients---owing to the choice of initial data~\eqref{eq:ICs}. 
To simplify the computation of Taylor coefficients and the subsequent application of asymptotic methods, it is advantageous to transform the time and fluid variables according to
\be \label{eqs:rescale}
  \boxed{ t^2 = \tau \,, \qquad   u(z,t) = U(z,\tau) \,, \qquad v(z,t) = t \,V(z,\tau) \,,
   \qquad  w(z,t) = t\, W(z,\tau) \,}\,, 
\ee
and perform the analysis in that rescaled system. Applying the transformation~\eqref{eqs:rescale} to~\eqref{eqs:HL}, we arrive at the rescaled HL model
\be 
  \partial_\tau U + \frac{1}{2} V \nabla_z U = 0 \,, \qquad
  \partial_\tau W + \frac{1}{2\tau} \left( W - \nabla_z U \right) = -\frac 1 2 V \nabla_z W\,, \qquad \nabla_z V = {\cal H}[W] \,.
  \label{eqs:HLrescaled}
\ee
These equations can be straightforwardly solved by the following Taylor-series {\it Ans\"atze},
\be \label{eqs:ansaetze}
   U(z,\tau) = \sum_{n=0}^\infty U_n(z)\,\tau^n \,, \qquad 
   V(z,\tau) = \sum_{n=0}^\infty V_n(z)\,\tau^n \,, \qquad
   W(z,\tau) = \sum_{n=0}^\infty W_n(z)\,\tau^n \,, 
\ee
which, by construction, possesses exclusively non-vanishing coefficients (except at special locations~$z$, see below). 
It is then elementary to derive all-order recursive relations for the  Taylor coefficients in Eq.~\eqref{eqs:ansaetze} (see e.g.\ \cite{Zheligovsky:2013eca,Rampf:2022apj, KolluruPandit2024}). Indeed, plugging~\eqref{eqs:ansaetze} into~\eqref{eqs:HLrescaled} and identifying the various powers in $\tau$, one finds~($n>0$)
\begin{subequations}  \label{eqs:eul_rec_hl1d}
\begin{align}
U_{n+1} &= \frac{-1}{2n+2} \sum_{l+m = n} V_l \,\partial_z U_m \,, \\
V_{n+1} &= \int \mathcal{H}\left[ W_{n+1} (z') \right] \dd z'\,, \\ 
W_{n+1} &= \frac{-1}{2n+3} \bigg({-}  \partial_z U_{n+1} +  \sum_{l+m = n} V_l \partial_z W_m  \bigg) \,.
\end{align}
\end{subequations}
For convenience, we provide the code implementing these (and the Lagrangian) recursive relations in App.\,\ref{app:recs}.
We remark that Ref.\,\cite{KolluruPandit2024} derived recursive relations for the HL model formulated in $t$ time, which are more general to ours, since their recursive relations apply also to asymmetric initial data; by contrast, our recursive relations are more suitable for investigating the asymptotic nature of the HL model based on symmetric initial data such as~\eqref{eq:ICs}.

Using the recursive solutions~\eqref{eqs:eul_rec_hl1d} (or, in mathematically equivalent form, those of Ref.\ \cite{KolluruPandit2024}), it is straightforward to generate the Taylor coefficients to fairly large orders. Specifically, we determine the first 36 $\tau$-time (72 $t$-time) Taylor coefficients and evaluate them {\it symbolically}, in contrast to Ref.\,\cite{KolluruPandit2024}, who sampled them instead on an equispaced mesh using a $2/3$-dealiased Fourier pseudo-spectral method with quadruple-precision arithmetic.

\begin{figure} 
\centering 
\includegraphics[width=0.99\textwidth]{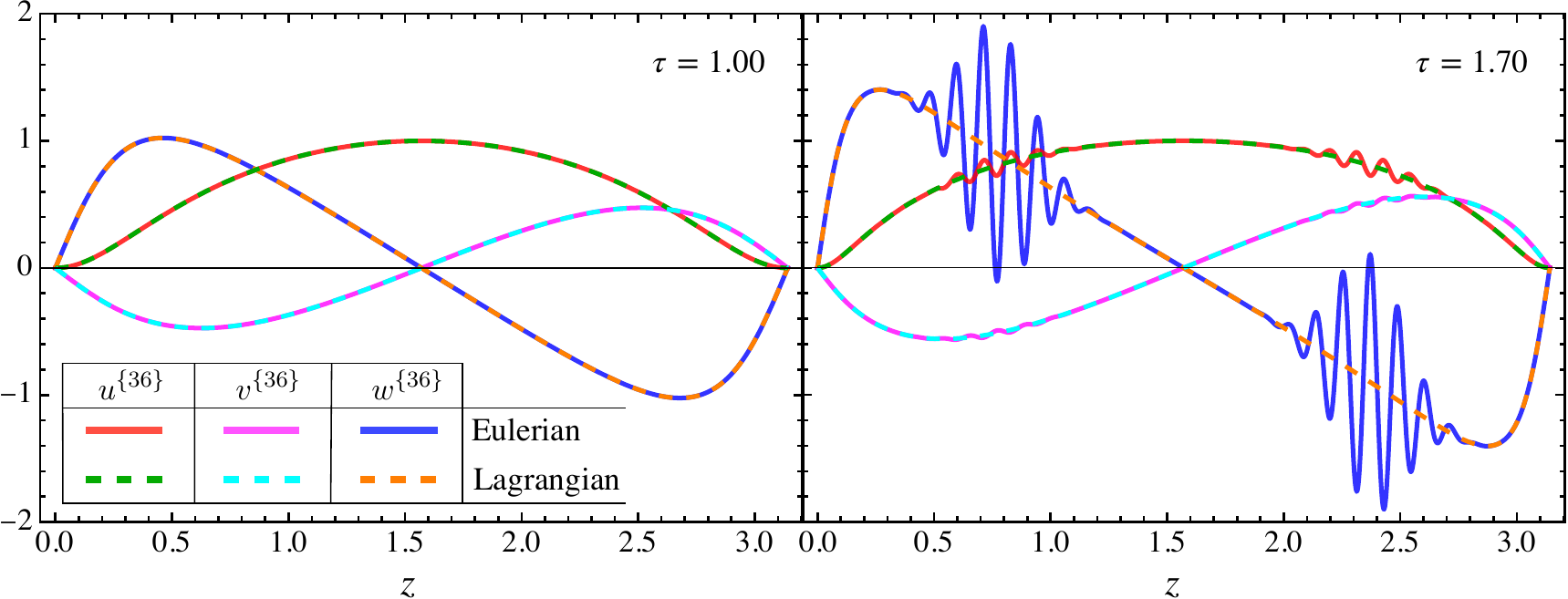}
\caption{Fluid variables $u,v,w$ in the HL model, represented by $\tau$-time series solutions formulated in Eulerian (solid lines) and Lagrangian coordinates (dashed lines; see Sec.\,\ref{sec:LagAnalysis}), where $\tau = t^2$. All Taylor series solutions are truncated at 36th order in $\tau$-time, and the Taylor coefficients are evaluated symbolically. Only half of the periodic domain in $z$ is shown; $u$ is mirror symmetric while $v,w$ are point symmetric due to the initial data~\eqref{eq:ICs}.
{\it Left panel:}  Fluid variables  evaluated at time $\tau=1$, where Eulerian and Lagrangian methods agree to better than $2 \times 10^{-9}$ precision (not shown). {\it Right panel:} Solutions at $\tau = 1.7$, where the Eulerian solutions display pathological behavior long before the time of the blow-up at $\tau_\star \simeq 4.483$ (see also Fig.\,\ref{fig:sing}), in contrast to the regular evolution retained in the Lagrangian solutions.} 
\label{fig:fluid-variables}
\end{figure}

In Fig.\,\ref{fig:fluid-variables}, we show the temporal evolution of the fluid variables $u = U$, $v= t V$ and $w = t W$, employing the Taylor-series solutions $F^{\{N\}} := \sum_{n=0}^N F_n \tau^n$ for $F= U,V,W$ up to truncation order $N =36$, shown respectively with red, magenta and blue solid lines. At time $\tau=1$ (left panel), the Eulerian solutions agree well with the series solutions formulated in Lagrangian coordinates (dashed lines in green, cyan and orange; see Sec.\,\ref{sec:LagAnalysis} for details). However, at time $\tau = 1.7$ shown in the right panel, the Eulerian solutions begin to deteriorate with the appearance of initially local, oscillatory structures, here centered about $z \simeq \pm\pi/4, \pm3\pi/4$ (only half of the periodic $z$-domain is shown). These resonant features, dubbed ``early-time tygers,'' have been observed for the case of inviscid Burgers' \cite{Rampf:2022apj} as well as for the HL model at hand \cite{KolluruPandit2024}. In the more wider fluid mechanical context, the general tyger phenomenon has recently received considerable attention (e.g.\ \cite{Ray2011, 2013NJPh...15k3025S, Podvigina2016, 2020PhRvR...2c3202M, KolluruBessePandit2024}.

\subsection{Asymptotic extrapolation in Eulerian coordinates}\label{sec:asyEuler}

Here we reiterate and slightly extend the analysis of Ref.\,\cite{KolluruPandit2024}. Specifically, we re-apply the Mercer--Roberts (MR) method to detect singular behavior in the HL model, and obtain MR estimates of the location and singularity exponent of the blow-up (see also App.\,\ref{app:MR} for yet another estimator for the blow-up time). In the complex-time vicinity near the blow-up, however, the MR method remains approximative, thus the resulting estimates should be interpreted with due care.

The MR analysis requires a real-valued model function $W^{\rm E}(\fett \tau)$, depending on complex time~$\fett \tau$, that (presumably) encapsulates the leading-order singular behavior of the vorticity~$W=w/t$ resulting from competing pairs of singularities in complex-time plane,
\be \label{eq:model}
   W^{\rm E}(\fett \tau) = \left( 1- \frac{\fett \tau}{\fett \tau_\star} \right)^\nu 
   +\left( 1-\frac{\fett \tau}{\fett {\bar \tau}_\star}  \right)^\nu, \quad 
   \fett \tau_\star \coloneqq R {\rm e}^{\ii \theta} \,, \quad
    R \in \mathbb{R}_+ \,, \quad \theta \in (-\pi,\pi] \,, \quad
    \nu \in  \mathbb{R} \setminus \mathbb{Z}_{\ge 0}\,,
\ee
and similarly for the other fluid variables. Here, $\fett \tau_\star$ and $\fett {\bar \tau}_\star$ denote the positions of complex-conjugated singularities with modulus~$R$ and angle~$\theta$, while $\nu$ is a real-valued singularity exponent; we emphasize that these unknowns (and thus $W^{\rm E}$) depend on Eulerian position~$z$, which we frequently suppress for notational simplicity. For $|\fett \tau| < R$, the model function~\eqref{eq:model} can be represented by a Taylor series $W^{\rm E}(\fett \tau)= \sum_{n=0}^\infty W^{\rm E}_n \fett \tau^n$ with Taylor coefficients~$W^{\rm E}_n = 2 (-1)^n \tbinom{\nu}{n} R^{-n} \cos(n \theta)$. By considering a suitable sequence $B_n$ constructed from $W^{\rm E}_n$, Mercer and Roberts~\cite{MercerRoberts} showed that the leading-order asymptotic behavior associated with pairs of complex-time singularities can be estimated as
\begin{align}
  B_n  &= \frac{1}{R} \left(1  - \frac{\nu +1}{n} + O(n^{-2}) \right)  \!, \quad\,\,
  \cos \theta_n = \cos \theta \bigg(   1+ \frac{\nu+1}{n^2} \left[  1 - \frac{\cos(2n-1)\theta}{\cos \theta} \right]  + O(n^{-3})  \bigg) , \label{eqs:MRextra}
\end{align}
where the estimators are 
\begin{align}
 B_n^2 &=  \frac{W_{n+1} W_{n-1} - W_n^2}{W_{n} W_{n-2} - W_{n-1}^2} \,, \qquad 
 \cos \theta_n = \frac 1 2 \left( \frac{W_{n-1}}{W_n} B_n   + \frac{W_{n+1}}{W_n } B_n^{-1} \right) \,.  \label{eq:MRestimators}
\end{align}
Here we have replaced the generic coefficients $W_n^{\rm E}$ with the time-Taylor coefficients of the HL vorticity~$W_n$ obtained from the recursive relations~\eqref{eqs:eul_rec_hl1d}. The MR analysis then rests on the assumption that, at sufficiently large order, the asymptotic behavior of $W_n$ is governed by the same singular structure as the model sequence $W_n^{\rm E}$. In practice, the unknown parameters $R$, $\theta$, and $\nu$ are extracted by plotting $B_n$ against $1/n$ [and $\cos\theta_n$ against $1/n^2$] and extrapolating to the limits $1/n\to0$ [$1/n^2\to0$], corresponding to $n\to\infty$; see App.\,\ref{app:MR} for details.

\begin{figure}
\centering 
\includegraphics[width=\textwidth]{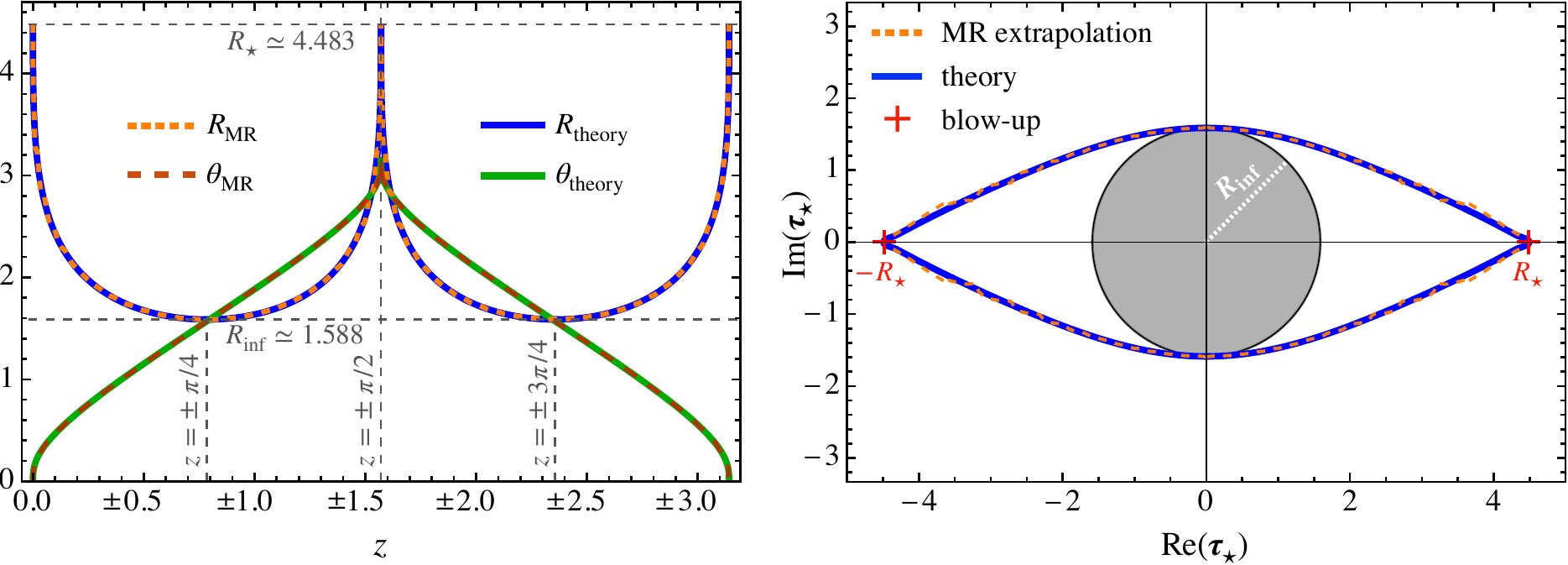}
\caption{{\it Left panel:} Modulus~$R$ and phase~$\theta$ of closest singularities~$\fett \tau_\star = R \exp(\ii \theta)$ as a function of~$z$, obtained from the Mercer--Roberts (MR) asymptotic extrapolation method (orange dot-dashed and red-dotted results), 
and from the singularity theory developed in Sec.\,\ref{sec:sing-theory} (blue and green solid lines). The minimal radius of convergence,  $R_{\rm inf} = R^{\MR}(z_{\rm ty}) \simeq 1.588$, occurs at $z_{\rm ty} = \pm \pi/4,\pm 3\pi/4$ with phase $\theta_{\MR}(z_{\rm ty}) \simeq \pi/2$
(accurate to 11 digits). 
These singularities lie on the imaginary axes, and trigger the early-time tygers seen in the Eulerian time-Taylor solutions (Fig.\,\ref{fig:fluid-variables}).
{\it Right~panel:} Convergence-limiting singularities in the complex-$\fett \tau$ plane, based on the MR method (orange dashed line) and on our theory (blue solid line). 
} \label{fig:sing}
\end{figure}

In the left panel of Fig.\,\ref{fig:sing}, we summarize the results from the MR method (dashed lines), using the $\tau$-time coefficients $W_n$ with $n=30-35$ as input for~\eqref{eqs:MRextra}; for convenience we also show our singularity theory results (solid lines) derived and discussed further below (see Sec.~\ref{sec:sing-theory}).
The MR methods reveals that the singularities closest to the origin are located at a distance (modulus) $R_{\rm inf} = R_{\MR}(z_{\rm ty}) \simeq 1.588$ and a phase $\theta(z_{\rm ty}) \simeq \pi/2$, and correspond to Eulerian positions $z = z_{\rm ty} = \pm \pi/4,\pm 3\pi/4$.
The nearest singularities are thus aligned with the imaginary time axis; this explains the birth of the early-time tygers at times $|\fett \tau| \gtrsim R_{\rm inf}$ observed in the right panel of Fig.\,\ref{fig:fluid-variables}. At the locations $z = z_\star = 0, \pm \pi/2, \pm \pi$, the MR method predicts real singularities  $R_{\MR} (z_\star)\simeq 4.486$
with a singularity exponent of $\nu_\star \simeq -3.06$, retrieved by considering the limits $z \to z_\star$ before evaluating~$B_n$. Near these singularities, the leading-order MR estimator~\eqref{eqs:MRextra} becomes largely approximative as higher-order terms in $1/n$ become essential for further analysis. See Ref.\,\cite{Rampf:2022apj} for a nonlinear extrapolation method developed for inviscid Burgers, which may partially remedy these limitations of the MR method. Despite its approximate nature, the MR results agree reasonably well with those obtained from the singularity theory developed in Sec.\,\ref{sec:sing-theory}.

In the right panel of Fig.\,\ref{fig:sing}, we present the MR results  (orange dashed line) in the complex-$\fett \tau$ plane, together with the disc of convergence (in gray) with radius $R_{\rm inf} \simeq 1.588$, which limits the validity regime of the Eulerian series solutions for the vorticity---but not the Lagrangian theory (blue solid line) which is valid within the complex disc $0 \leq |\fett \tau| < R_\star \simeq 4.483$. 
Relative to the theoretical prediction, the MR extrapolation determines $R_{\rm inf}$ with an accuracy of $0.013\%$ at the present truncation order ($n=36$).
For singularities close to---but off---the real time axis, the MR extrapolation results show mild oscillatory behavior relative to the theoretical results, which is a feature already observed (in a milder form) for inviscid Burgers \cite{Rampf:2022apj}. In contrast, both MR extrapolation and Lagrangian singularity theory predict real-valued singularities at $R = \pm R_\star$ with   $R_\star \simeq 4.483$ and $R_{\MR} (z_\star)\simeq 4.486$, differing from each other by about $0.070\%$. Lastly, regarding the validity regime of the Eulerian solutions, we remark that an analytic continuation (\`a la Weierstrass)  along the real time could be feasible by e.g.\ performing consecutive expansions at fresh expansion points $\tau = \tau_i$ with $\tau_i < R_{{\rm inf}, i-1}$, where $R_{{\rm inf}, i-1}$ denotes the convergence radius of the previous time step $(i-1)$; here, $i=1,2,\ldots$ with $R_{{\rm inf}, 0} = R_{\rm inf}$ and $\tau_1 = 0$ (see Ref.\,\cite{Podvigina2016} for related avenues).

We conclude that the MR extrapolation method leads to fairly reliable results, except for singularities located close to the real time axis---as already observed in previous works~\cite{Rampf:2022apj,KolluruPandit2024}. Ultimately, however, the MR method comes with measurement uncertainties and is phenomenological in nature. To get deeper insights into the skeleton of the HL model, we develop a Lagrangian singularity theory in Sec.\,\ref{sec:sing-theory}; we discuss requisite preliminary considerations in Secs.\,\ref{sec:LagAnalysis}--\ref{sec:UV} below.

\section{Hou--Lou model in Lagrangian space} \label{sec:LagAnalysis}

Instead of considering the HL equations~\eqref{eqs:HL} in a fixed Eulerian frame of reference, we now formulate and analyze them using Lagrangian coordinates. The corresponding equations are recast in Lagrangian form and solved with a Taylor-series {\it Ans\"{a}tze} in Sec.\,\ref{sec:HLlagspace}. In Sec.\,\ref{sec:asyLag}, we perform an asymptotic analysis of the Lagrangian solution. While the Lagrangian solution blows up, similarly as in the Eulerian case, we find that its analytic structure is significantly simplified.

\subsection{Formulation in Lagrangian space and series representation}\label{sec:HLlagspace}

The HL model~\eqref{eqs:HL} can be recast using characteristic (Lagrangian) coordinates $a$. For this we introduce the characteristic map $a \mapsto {\cal Z}(a,t)$ from initial position~$a$ at time $t=0$ to current position~${\cal Z}(a,t)$ at time~$t$. The map ${\cal Z}(a,t)$ encodes the trajectories of fluid particles advected by~$v$ along the (Eulerian) $z$-direction, and satisfies the characteristic equation~$v = \dd {\cal Z}/\dd t$, where $\dd/\dd t = \partial_t +v \nabla_z\,$.
The Jacobian  
\be  \label{eq:Jac}
 {\cal J} := \nabla_a {\cal Z}
\ee
measures the separation of fluid particles along the $z$ direction and thus, {\it is not fixed to unity}---as it is e.g.\ the case for the Jacobian (determinant) of the Lagrangian mapping for the incompressible Euler equation (see e.g.\ \cite{Zheligovsky:2013eca}). Instead, the time evolution of the present Jacobian is governed by $\dd {\cal J}/ \dd t = {\cal J} \partial_z v$ which, upon using the Biot--Savart law in~\eqref{eqs:HL}, can be written as $\dd {\cal J}/ \dd t ={\cal J} {\cal H}[w]$. Actually, as we will shortly see, multiple fluid particles approach the same Eulerian position ${\cal Z}$ as $t \to t_\star$, causing the Jacobian~\eqref{eq:Jac} to vanish at $a=a_\star$---this is the driving mechanism of the blow-up in the HL model.

With this prelude, we can rewrite Eqs.\,\eqref{eqs:HL} in the compact form 
\begin{align} \label{eqs:HLlagInterm}
      \frac{\dd}{\dd t} u = 0 \,, \qquad   \frac{\dd}{\dd t}{\cal Z} &= v \,,  \qquad \frac{\dd}{\dd t} w = \nabla_z u \,, \qquad  
  \frac{\dd}{\dd t} {\cal J} = {\cal J} {\cal H}[w] \,,
\end{align}
where the remaining Eulerian gradient $\nabla_z $ is transformed to Lagrangian space below. These equations are supplemented with the initial conditions
\be \label{eqs:ICsLag}
  u(t=0)= \sin^2(a) \,, \qquad w(t=0) =0 \,.
\ee
Interestingly, the Lagrangian equation $\dd u/\dd t=0$ together with the first initial condition~\eqref{eqs:ICsLag} immediately implies the trivial solution
\be
   u({\cal Z}(a,t)) = \sin^2 a \,,
\ee
meaning that the variable $u$ is a temporal constant along fluid characteristics. Using this, together with the rescaling of time $\tau =t^2$ and fluid variables $u={\cal U}$, $v = {\cal V} t$, $w = {\cal W} t$, we arrive at the rescaled HL system in Lagrangian space,
\begin{align} \label{eqs:HL-Lag}
   \boxed{  {\cal U} =  \sin^2 a \,, \qquad    {\cal V} =2 {\cal Z}' \,, \qquad 
     2 \tau  {\cal W}' + {\cal W}   = \frac{\sin (2a)}{{\cal J}} \,, \qquad 
    {\cal J}' = \frac 1 2 {\cal J} {\cal H}[{\cal W}] \,} \,,
\end{align}
where ${\cal Z}' = \partial_\tau {\cal Z}$ etc., and all fluid variables are functions of ${\cal Z}(a,\tau)$.

To solve Eqs.\,\eqref{eqs:HL-Lag} we impose a Taylor-series {\it Ans\"{a}tze} for the so-called Lagrangian displacement field $\mPsi = {\cal Z} - a$,
\be \label{eq:psi}
   \mPsi(a,\tau) = \sum_{n=1}^\infty \mPsi_n(a)\,\tau^n \,,
\ee
where the $\mPsi_n$ are the $\tau$-time Taylor coefficients of the Lagrangian displacement field. To retrieve the other fluid variables in terms of~$\mPsi$, one could plug~\eqref{eq:psi} into~\eqref{eqs:HL-Lag} and identify the relationship between the various $\tau$-time coefficients, in a similar way as for the Eulerian case in Sec.\,\ref{sec:EulerianSeries}. In practice, however, such avenues become rapidly cumbersome because, to our knowledge, no closed-form expression of the Hilbert transform is available in Lagrangian coordinates.

Instead, we determine the $\mPsi_n$ recursively---and in a computationally cheap manner---by exploiting the fact that we have already determined the Eulerian solutions in Sec.~\ref{sec:EulerianSeries}. For this, we consider the Fourier transform of the Eulerian variable $V$, 
\be \label{eq:tildeV}
  \tilde V(k,\tau) := \int V(z,\tau) \,{\rm e}^{- \ii k z} \dd z
    = 2\int {\cal Z}'(a,\tau)\, {\rm e}^{- \ii k {\cal Z}(a,\tau)} {\cal J}(a,\tau) \,\dd a \,.
\ee
Here, we have used $\dd z = {\cal J} \dd a$ and the definition $V(a,\tau) = 2 \partial_\tau {\cal Z}(a,\tau)$. Now, using the Fourier-series representation $\tilde V(k,\tau) = \sum_{n=0}^\infty \tilde V_n(k)\,\tau^n$ with coefficient $\tilde V_n(k) = \int V_n(z) \exp(-\ii k z)\,\dd z$ in the first expression of~\eqref{eq:tildeV} and expanding the last expression of~\eqref{eq:tildeV} about $\tau=0$, we obtain an equation that contains infinite $\tau$-time series. Matching equal powers in $\tau^n$, we arrive at the following recursive relations for the displacement coefficients ($n\geq 0$)
\be 
    \boxed{ (n+1) {\cal F}_{a\to k}[\mPsi_{n+1}] = \frac{\tilde V_n(k)}{2}
      -   \sum_{\substack{p+q\\=\,n+1}} p {\cal F}_{a\to k}\left[\mPsi_p \left( \nabla_a \mPsi_q + E_q \right)\right] 
      -  \sum_{\substack{p+q+r\\=\,n+1}}^{\substack{\phantom{=}\\\phantom{=}}} p  {\cal F}_{a\to k}\left[ \mPsi_p E_q \nabla_a \mPsi_r  \right] \,
      } 
      \,. \label{eq:psi-rec}
\ee
Here, ${\cal F}_{a\to k}[\mPsi] := \int \mPsi(a) \exp(-\ii k a)\,\dd a$ denotes the Fourier transform w.r.t.\ the Lagrangian coordinate~$a$, and we have defined $E(a)=\exp(-\ii k \mPsi(a)) = 1+  \sum_{n=1}^\infty E_n \tau^n$, where $E_n$ is the $n$th-order Taylor coefficient of $E(a)$ about $\tau=0$. Equipped with $V_n$ retrieved from~\eqref{eqs:eul_rec_hl1d}, we can then efficiently generate the displacement coefficients to arbitrary order. For convenience, we provide Mathematica codes that implements both the Lagrangian and Eulerian recursive relations in App.\,\ref{app:recs}.

Once the displacement is known up to a given truncation order, one can easily determine the corresponding solutions for the Lagrangian representations 
\be \label{eqs:VandW}
  {\cal V}(a,\tau) = \sum_{n=0}^\infty {\cal V}_n(a) \tau^n \,, \qquad 
  {\cal W}(a,\tau) = \sum_{n=0}^\infty {\cal W}_n(a) \tau^n
\ee
from the corresponding equations in~\eqref{eqs:HL-Lag}. The first few terms in the resulting series read
\begin{align}
  \mPsi(a,\tau) &= -\tau \frac{\sin a \cos a}{2} +\tau^2 \frac{\sin (4 a)}{96} + \tau^3  \frac{24 \sin(2 a) - \sin(6 a)}{4320} + O(\tau^4) , \label{eq:psi-sol}\\
  {\cal V}(a,\tau) &=  -\frac{\sin(2a)}{4} + \tau \frac{\sin(4a)}{48} 
    - \tau^2 \frac{\sin(6a)-24 \sin (2a)}{1440} + \tau^3 \frac{3 \sin(8a)-80 \sin(4a)}{241920} + O(\tau^4),\\
  {\cal W}(a,\tau) &= \sin (2 a) + \tau \frac{\sin(4a)}{12}  + \tau^2 \frac{2 \sin (2 a)+\sin (6 a)}{120}
   + \tau^3 \frac{72 \sin (4 a)+17 \sin (8 a)}{20160} + O(\tau^4). \label{eq:W} 
\end{align}
In Fig.\,\ref{fig:fluid-variables} we show the Lagrangian predictions for $u(z),v(z),w(z)$ (various dashed lines), by plotting ${\cal U}(a),\tau^{1/2}{\cal V}(a), \tau^{1/2}{\cal W}(a)$ parametrically over ${\cal Z}(a)$, truncated to the 36th order in $\tau$-time. The Lagrangian predictions remain regular at the considered times $\tau=1.0, 1.7$ and are entirely unaffected by the singularities situated at~$R_{\rm inf} \simeq 1.588$ where the Eulerian series solutions cease to be meaningful.

\begin{figure} 
\centering 
\includegraphics[width=0.98\textwidth]{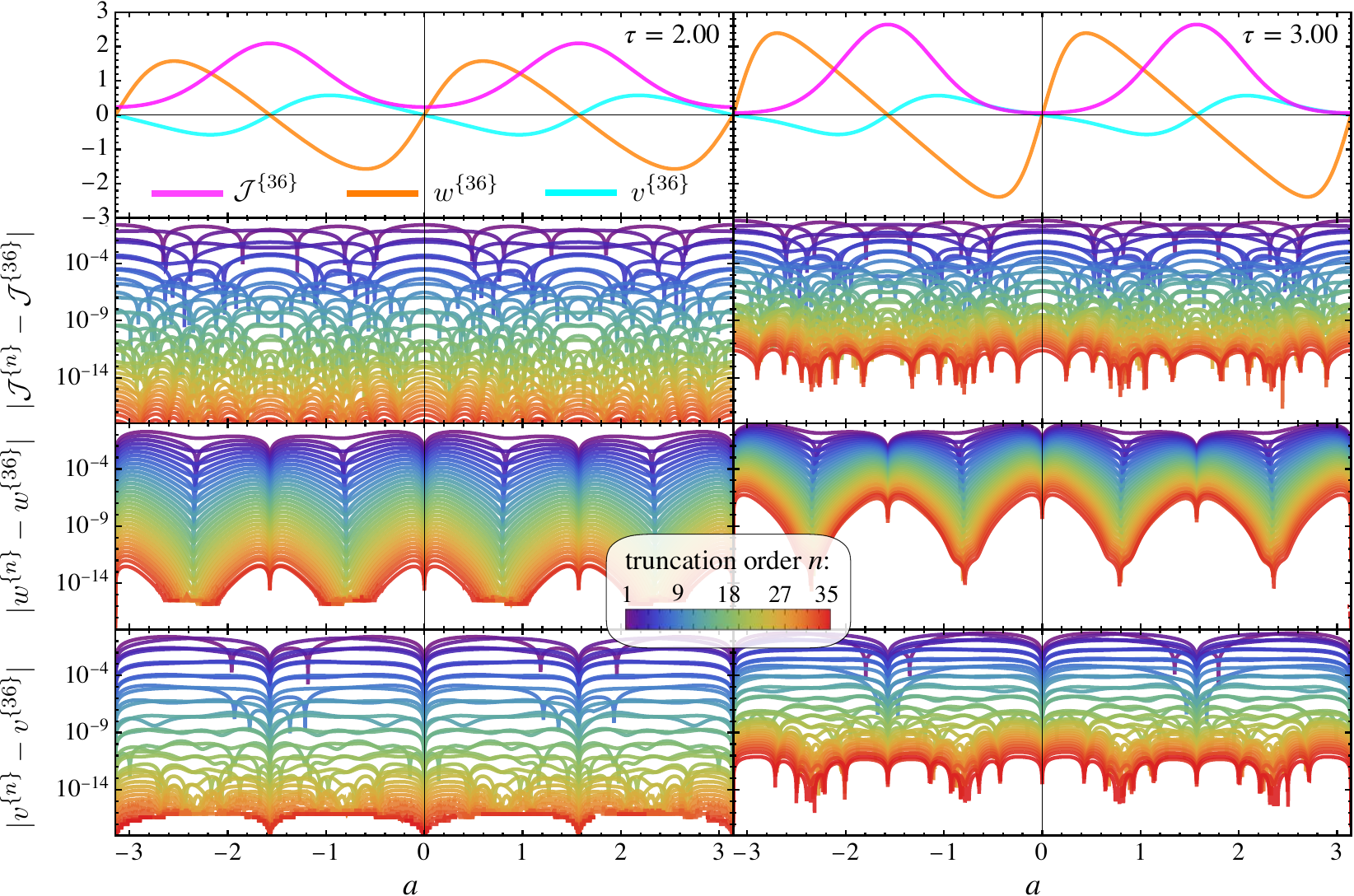}
\caption{Jacobian~${\cal J}=\nabla_a Z$ and fluid variables~$w = \tau^{1/2}{\cal W}$ 
and~$v = \tau^{1/2}{\cal V}$ as a function of Lagrangian coordinate~$a$, truncated at various orders $1 \leq n \leq 36$ (indicated by ``$\{n\}$''), evaluated at times $\tau =2$ (left panels) and $\tau=3$ (right panels). 
The top panels show the Lagrangian predictions at maximal truncation order, while the subsequent panels show their absolute differences w.r.t\ solutions at lower truncation orders~$1\leq n <36$. 
} \label{fig:lag-fluid-plots}
\end{figure}

To further elucidate the improved performance of the Lagrangian formulation, we show in Fig.\,\ref{fig:lag-fluid-plots} the Lagrangian fluid variables as a function of~$a$ for significantly later times, $\tau=2$ (left panels) and $\tau=3$ (right panels). In the top row, we show the Lagrangian $\tau$-time series solutions of the fluid variables at the maximal truncation order $N=36$, while the various sub-panels show the absolute difference of the series solutions as a function of truncation order~$n$ where $0< n < N$. The series solutions of all fluid variables appear to converge, with the expected trend that truncation errors are generally smaller at earlier times. Additionally, we observe a spatial dependence on the speed of convergence, with the vorticity $w^{\{ n\}}$  representing the most stringent trend: at high orders, the truncation errors are largest near the locations $a = 0, \pm \pi/2, \pm \pi$, although we note that the vorticity goes through zero crossings at precisely these locations. Moreover, as we explain below, the slow convergence speed near these locations is driven by singular behavior.

Finally, observe that ${\cal J}$ is strictly positive for the times considered in Fig.\,\ref{fig:lag-fluid-plots}, but, as we will see shortly, vanishes when $\tau \to \tau_\star$ at locations $a = a_\star = 0,\pm \pi$. At this instant, the separation of certain fluid particles along the $z$ direction thus collapses to zero, ultimately causing the mapping $a \to {\cal Z}$ to lose monotonicity.

\subsection{Asymptotic analysis in Lagrangian space}\label{sec:asyLag}

Based on the above findings summarized in Figs.\,\ref{fig:fluid-variables} and~\ref{fig:lag-fluid-plots}, it appears that the Lagrangian solutions are not limited by the same complex-time singularities as in the Eulerian case. This observation suggests that the analytic structure of the Lagrangian solutions may be simpler. To test this hypothesis, let us consider a standard Domb--Sykes analysis \cite{DombSykes1957,vanDyke1974} where the model function involves a single non-analytic term, specifically
\be \label{eq:ModelL}
  {\cal W}^{\rm L}(\fett \tau) = \left(1 - \frac{\fett \tau}{\tau_\star^{\smallL}} \right)^{\nu_{\mskip1mu\rm L}} \,, \qquad \text{with}\quad \tau_\star^{\smallL} = \tau_\star^{\smallL}(a) \ \in \ \mathbb{R}\,, \quad \nuL = \nuL(a) \ \in \ \mathbb{R} \setminus \mathbb{Z}_{\ge 0} \,,
\ee
for the Lagrangian vorticity. We remark that similar considerations also apply to the other fluid variables, however based on Fig.\,\ref{fig:lag-fluid-plots}, it appears that the vorticity primarily exhibits singular behavior, and this is why we focus the following analysis on this variable. Returning to~\eqref{eq:ModelL}, for $|\fett \tau|< \tau_\star^{\mskip1mu\rm L}$, the model function can be represented by the expansion
\be 
  {\cal W}^{\rm L}= \sum_{n=0}^\infty {\cal W}_n^{\rm L} \fett \tau^n \,, \qquad {\cal W}_n^{\rm L} = \binom{\nuL}{n} [-\tau_\star^{\smallL}]^{-n} \,,
\ee
where we employ the generalized binomial coefficient $\binom{\nuL}{n}$. Taking subsequent ratios of coefficients ${\cal W}_n^{\rm L}/{\cal W}_{n-1}^{\rm L}$, and replacing the model coefficients ${\cal W}_n^{\rm L}$ with the coefficients of the Lagrangian vorticity ${\cal W}_n$, one finds exactly 
\be \label{eq:Dn}
   D_n := \frac{{\cal W}_n}{{\cal W}_{n-1}} = \frac{1}{\tau_\star^{\smallL}} \left(  1 - [1+\nuL]\,\frac{1}{n}\right) \,.
\ee
In 
\begin{figure}
 \centering 
\includegraphics[width=0.99\textwidth]{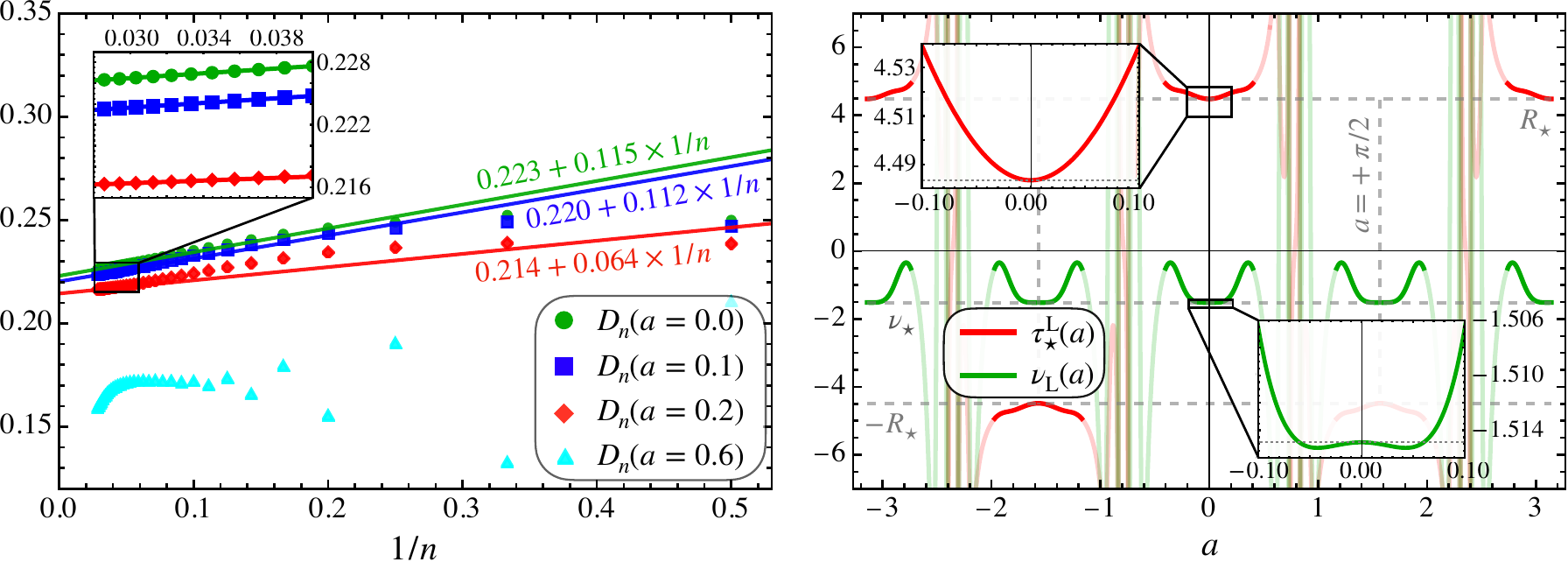}

\caption{{\it Left panel:}  Domb--Sykes  estimator~$D_n(a)$ [Eq.\,\ref{eq:Dn}] for the $\tau$-time Taylor coefficients of the Lagrangian vorticity up to order $N=36$. We evaluate $D_n(a)$ at four exemplary locations $a= 0,0.1,0.2,0.6$ (respectively shown in green, blue, red and cyan). For all locations except the last, the $D_n$ settle into a linear behavior for orders $n \gtrsim 30$ justifying linear extrapolation, thereby revealing the leading-order asymptotics of the Lagrangian vorticity.
{\it Right panel:}  Summary of results for the location of the temporal singularity $\tau_\star^{\smallL}$ (red) and singularity exponent $\nuL$ (green; see Eq.\,\ref{eq:ModelL}). Locations where the Domb--Sykes test is non-conclusive are drawn with faint lines.
} \label{fig:DS-Rofa}
\end{figure}
the left panel of Fig.\,\ref{fig:DS-Rofa}, we evaluate $D_n$ for $1 < n \leq 36$ at four exemplary locations $a=0.0,0.1,0.2, 0.6$. At all these locations---except at $a=0.6$ (see further below)---the $D_n$ ratios settle into a linear behavior for sufficiently large~$n$, suggesting that d'Alembert's limit in the ratio test
\be
   \frac{1}{R} = \lim_{n \to \infty} \left| \frac{{\cal W}_n}{{\cal W}_{n-1}} \right| 
\ee
exists, where $R$ is the radius of convergence. Performing a least-square fit for $D_n$ versus $1/n$ between orders $30 \leq n \leq 35$ allows us to estimate $\tau_\star^{\smallL}(a)$ (on $R$) and the singularity exponent~$\nuL(a)$ in~\eqref{eq:Dn}, in the limit of $n \to \infty$. At off-grid locations such as at $a=0.6$ (cyan/triangle plot markers in left panel), however, the $D_n$ do not settle into a regular behavior, at least not within the Taylor coefficients that we have at hand, therefore we do not take them into consideration for the asymptotic analysis (see below for related comments).

In the right panel of Fig.\,\ref{fig:DS-Rofa}, we show the resulting estimates of $\tau_\star^{\smallL}(a)$ and $\nuL(a)$ (red and green, respectively); faint lines indicate estimates where the Domb--Sykes analysis was inconclusive. Importantly, we detect real singularities, with those situated closest to the origin being
\be \label{eq:Rstar}
  \boxed{ R_\star := \tau_\star^{\smallL}(a_\star^+) \simeq 4.48339\,,
 \quad
   -R_\star = \tau_\star^{\smallL}(a_\star^-) \simeq - 4.48339\,,
   \quad \text{with~~} \nu_\star := \nu_{\smallL}(a_\star^\pm) \simeq -1.51487 \,
   }\,,
\ee
where $a_\star^+ = 0, \pm \pi$ and $a_\star^- = \pm \pi/2$. We thus conclude that the Lagrangian vorticity blows up at time~$\tau = R_\star$, with leading-order asymptotics (asymp) being encapsulated by
\be \label{eq:Winf}
 {\cal W}^{\rm asymp} (a) = C \big( \tau_\star^{\smallL} -\tau \big)^{\nuL} \,, \qquad \text{for}~ \left| a - a_\star^\pm \right| \lesssim 0.4 \,, 
\ee
where $C$ is a constant, and we remind the reader that both $\tau_\star^{\smallL}$  and $\nuL$ are (weakly) dependent on location in the neighborhood $a \simeq a_\star^\pm$. Note that the safe evaluation of the ratios of trigonometric functions appearing in $D_n(a)$ at $a_\star^+$ requires the limit $a \to a_\star^+$. Regarding the off-grid locations, it is plausible that the asymptotic modeling could be refined by taking the competing singularities along the positive and negative time axis into account, which would likely alleviate the ``interference patterns'' observed in the right panel of Fig.\,\ref{fig:DS-Rofa}. However, since only the singularities closest to the origin are convergence-limiting, and since these singularities are located on the real time axis, we consider the above analysis sufficient for our purposes.

We emphasize that, although the nearest  singularities are located precisely at $a= a_\star^+$, the series solution for the vorticity is {\it exactly zero} at these locations (cf.\  Eq.\,\ref{eq:W}, see also Sec.\,\ref{sec:UV} below). This is despite the fact that  $\lim_{a\to a_\star^+}D_n(a)$ exists and is nonzero. Moreover, the Domb--Sykes estimator near the blow-up locations is well converged; this should be contrasted to the Eulerian results relying on the MR estimator which have significant precision errors near the blow-up (Fig.\,\ref{fig:sing}).

It is intriguing to compare the Lagrangian results with those obtained in Eulerian coordinates. Specifically, based on the MR estimator evaluated at the Eulerian locations~$z=0,\pm \pi/2, \pm \pi/2$, we report in Sec.\,\ref{sec:asyEuler} a real-valued temporal singularity with modulus~$R_\star^{\MR} \simeq  4.4865$, which differs from the present estimate $R_\star \simeq 4.48339$ to~$0.069\%$. Actually, we have obtained yet another Eulerian estimate on the real singularity, namely $R_\star^{\DSE} \simeq 4.4842$ by performing the Domb--Sykes (DS) test for the Eulerian vorticity precisely at such blow-up locations (see App.\,\ref{app:MR}), which agrees with $R_\star$ to better than $0.018\%$.

Thus, we conclude that both Eulerian and Lagrangian solutions are limited by the same real singularities as expected, and that the small discrepancies between the three estimates most likely stem from residual extrapolation errors. These inaccuracies could be reduced by relying on higher-order Taylor coefficients.
For comparison, in the case of the 1D inviscid Burgers with smooth initial data where the blow-up time (pre-shock) can be determined exactly, the blow-up time was estimated with the MR method with a precision of $0.0061\%$ \cite{Rampf:2022apj}, albeit using 70 rather than the present 35~Taylor coefficients. If one repeats the analysis of Ref.\,\cite{Rampf:2022apj} for inviscid Burgers but restricts the evaluation of the MR estimator to orders $30$--$35$ (as done here for HL), the resulting blow-up time is accurate to only $0.012\%$, i.e., the error roughly doubles. Thus, based on these external considerations, one may expect that the errors in the extrapolation methods for HL could be reduced by about a factor of two when doubling the number of series coefficients from 35 to 70. Such a doubling, however, would come with an exponential increase in computational time, given that the series coefficients become exponentially larger at high orders, and this is why we defer such avenues to future work.

In the following section, we propose a simple model that incorporates the leading-order asymptotics obtained above. In Sec.\,\ref{sec:BKM} we then use this asymptotically completed model as a surrogate simulation output and test the extent to which the local asymptotics can be recovered from the global Beale--Kato--Majda criterion. See Sec.\,\ref{sec:sing-theory} for the construction of the singularity theory.

\section{Asymptotic completion (AC)}\label{sec:UV}

While the series representation for the Lagrangian vorticity~\eqref{eq:W} appears to converge within the complex-time disc of radius~$R_\star \simeq 4.48339$, its convergence is rather slow (Fig.\,\ref{fig:lag-fluid-plots}). 
Here, we apply an approach that exploits the asymptotic insights from Section~\ref{sec:asyLag} to achieve substantially faster convergence than the raw series expansion. This approach, dubbed the AC method hereafter, has recently been applied to cosmological fluids under the name UV method~\cite{Rampf:2022eiu,Rampf:2023fzq} (we adopt the term AC to avoid confusion with ultraviolet terminology); see Ref.\,\cite{vanDyke1974} for historical context.

In Sec.\,\ref{sec:asyLag}, we have deduced that the leading-order asymptotic behavior of the Lagrangian vorticity is encapsulated by ${\cal W}^{\rm asymp}= C (\tau_\star^{\smallL} -\tau)^{\nuL}$ at locations near the blow-up~$a_\star^\pm=  0, \pm \pi/2, \pm \pi$. The AC method suggests that the truncated Taylor series of ${\cal W}$ can be asymptotically completed by treating ${\cal W}^{\rm asymp}$ as a remainder of the infinite series. The asymptotically completed prediction of ${\cal W}$, dubbed AC vorticity or ${\cal W}^{\{n \rm AC\}}$ in the following, is of the following form
\be
  {\cal W}^{\{n \rm AC\}} := \sum_{s=0}^{n-1} {\cal W}_s \tau^s  + {\cal W}^{\rm asymp} - {\cal W}^{\{{\rm asymp}, n-1\}} \,, \qquad a \simeq a_\star^\pm \,.
\ee
Here, ${\cal W}_s$ are the Lagrangian coefficients~\eqref{eq:W}, while
${\cal W}^{\{{\rm asymp}, n-1\}}$ is the Taylor series of ${\cal W}^{\rm asymp}= C (\tau_\star^{\smallL} -\tau)^{\nuL}$ about $\tau=0$ (see below) truncated up to order $n-1$; this term is needed to avoid double counting of low-order coefficients up to truncation order $n-1$.

Within the AC method, the approximate results for $\tau_\star^{\smallL}$ and $\nu_{\smallL}$---obtained in the present case from the Domb--Sykes method (Fig.\,\ref{fig:DS-Rofa})---serve as the sole input. The only remaining task is to fix the constant $C$ appearing in Eq.\,\eqref{eq:Winf}, which we do so by demanding that the series coefficient of the Taylor series for ${\cal W}^{\rm asymp}$ is identical with the series coefficient for ${\cal W}$ at truncation order $n$. This matching then leads directly to the AC vorticity
\be \label{eq:WnUV}
  {\cal W}^{\{n \rm AC\}} = \sum_{s=0}^{n-1} {\cal W}_s \tau^s + \frac{{\cal W}_n}{c_n} \left[ \left(1- \frac{\tau}{\tau_\star^{\nuL}} \right)^{\nuL}   - \sum_{k=0}^{n-1} c_k \tau^k \right] \,, \qquad a \simeq a_\star^\pm \,,
\ee
where $c_n = \binom{\nu_{\smallL}}{n} [-\tau_\star^{\smallL}]^{-n}$ involves a generalized binomial coefficient. For example, at third order the AC vorticity reads ($a \simeq a_\star^\pm$)
\begin{align} \label{eq:W3UV}
   {\cal W}^{\{3 \rm AC\}} &= \sin(2 a) + \frac{\tau}{6} \cos(2 a) \sin(2 a) + 
 \frac{\tau^2}{120} \left[ 2 \sin(2 a) + \sin(6 a) \right] + \frac{\tau_\star^{\smallL} \Gamma(-\nuL)}{6720 \Gamma(3 - {\nuL})} \nonumber \\
&\quad \times   \left(  
   2 \tau_\star^2 \left[  \left( 1 - \frac{\tau}{\tau_\star^{\smallL}} \right)^{\nuL} - 1 \right] + 
   {\nuL} \tau (\tau + 2\tau_\star^{\smallL}) -\nuL^2 \tau^2 \right)  \left[72 \sin(4 a) + 
   17 \sin(8 a) \right] \,.
\end{align}
By construction, ${\cal W}^{\{n \rm AC\}}$ possesses the same leading-order asymptotic behavior for any truncation order~$n$. However, $n$ should be chosen sufficiently large to ensure competitive performance at early times, compared with the raw series evaluated at maximal truncation order.

\begin{figure}
 \centering 
\includegraphics[width=0.99\textwidth]{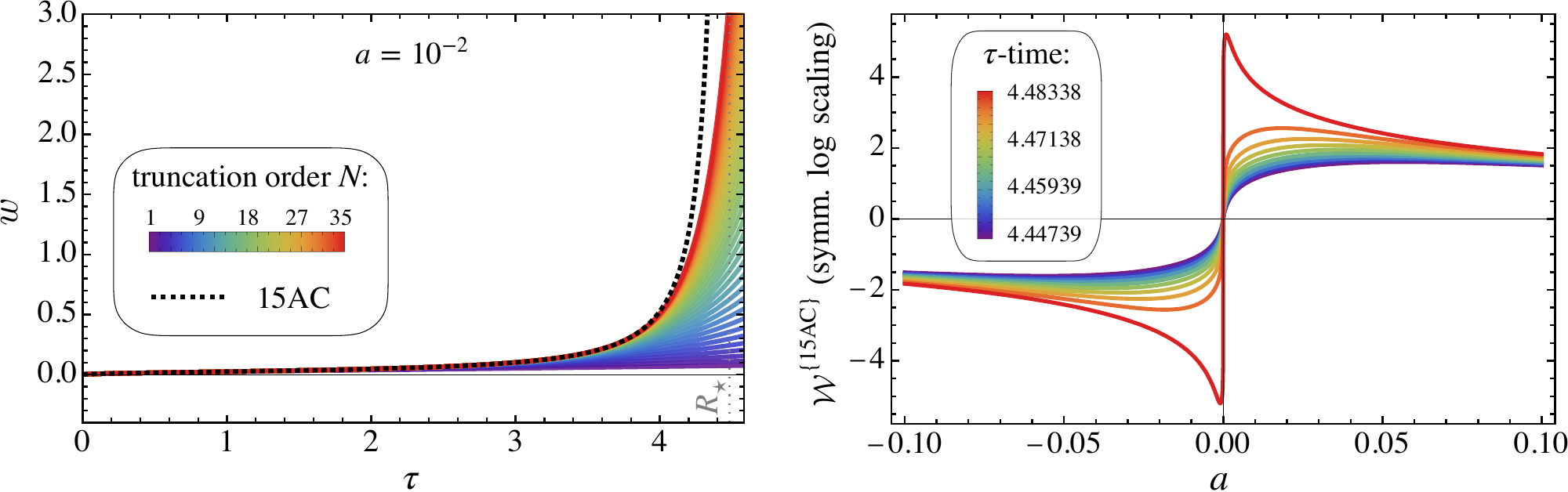}

\caption{{\it Left panel:} Lagrangian vorticity~$w = \tau^{1/2} {\cal W}$ as a function of~$\tau$ using ${\cal W}^{\{15 \rm AC\}}$ from the AC method~(black dotted line, Eq.\,\ref{eq:WnUV}) compared against truncated raw series solutions (various colors indicate truncation orders, Eq.\,\ref{eq:W}); all predictions are evaluated at $a=10^{-2}$.
{\it Right panel:} ${\cal W}^{\{15 \rm AC\}}$ in symmetric-log scaling as a function of $a$ in the neighborhood of the blow-up location~$a=a_\star =0$. The various colors (blue to red) indicate evaluation times approaching the blow-up time $\tau =R_\star \simeq 4.48339$. See also Fig.\,\ref{fig:BKM} showing the temporal evolution of the maxima/minima, relevant for applying global blow-up criteria.
} \label{fig:UV-vorticity}
\end{figure}

In the left panel of Fig.\,\ref{fig:UV-vorticity}, we show the temporal evolution of various approximations for the Lagrangian vorticity~$w =\tau^{1/2}{\cal W}$, in particular exploiting the AC model~${\cal W}^{\{15 \rm AC\}}$ (black dotted line) as well as various truncated series solutions for~${\cal W}$ (solid lines, color-coded as indicated in the legend). All functions are evaluated slightly off-grid from the blow-up location, namely at $a = 10^{-2}$, since the vorticity coefficients are identically zero at~$a=0$. For the AC method, we use the inputs $\tau_\star^{\smallL}(10^{-2}) \simeq 4.484 > R_\star \simeq 4.483$ and $\nuL(10^{-2}) \simeq \nu_\star \simeq -1.515$. It is seen that the AC vorticity agrees well with high-order raw series solutions at earlier times, while signaling a blow-up near the locations $a_\star^\pm$ as the system approaches the critical time, as expected.

In the right panel of Fig.\,\ref{fig:UV-vorticity}, we show the AC vorticity for $-0.1 \leq a \leq 0.1$---in the neighborhood of the blow-up location $a_\star =0$---evaluated at times very close to the blow-up occurring at~$\tau =R_\star \simeq 4.48339$ (blue to red color code indicates later times). Specifically, we show the symmetric log of the AC vorticity, defined with ${\rm symlog} ({\cal W}^{\{15 \rm AC\}}) = {\rm sign}( {\cal W}^{\{15 \rm AC\}})\log_{10}(1+ |{\cal W}^{\{15 \rm AC\}}|)$. Note that within the AC model, we do take the spatial dependencies of $\tau_\star^{\smallL}$ and $\nuL$ into account (for the input, see right panel in Fig.\,\ref{fig:DS-Rofa}). It is seen that, as $\tau$ approaches $R_\star$, the vorticity profile exhibits a clear ``squashing'' behavior: two nearby peaks form on either side of the origin---one for $a<0$ and one for $a>0$---while the vorticity at $a=0$ remains identically zero for $\tau < R_\star$. Evaluating the AC vorticity~\eqref{eq:W3UV} in the limit $\tau \to R_\star$, the most singular term is $\propto a (\tau_\star^{\smallL} -\tau)^{\nuL} + O(a^3)$ which is discontinuous around $a=0$. Since the Lagrangian point $a=0$ maps to ${\cal Z}=0$ at all times [$\mPsi(a=0)=0$ for any $\tau < \infty$; see Eq.\,\ref{eq:psi-sol}], the Eulerian vorticity similarly shows a squashing behavior for $\tau < R_\star$ and becomes discontinuous for~$\tau \to R_\star$ near~$z=0$.

It is enlightening to employ the AC vorticity~\eqref{eq:WnUV} to retrieve an AC prediction for the Jacobian. This task can be done by replacing ${\cal W} \to {\cal W}^{\{n \rm AC\}}$ in ${\cal W} +2 \tau  \partial_\tau {\cal W}  = \sin (2a)/{\cal J}$ [Eq.~\eqref{eqs:HL-Lag}], followed by an inversion of that relation, leading to the ``AC Jacobian''
\be \label{eq:jacUV}
 {\cal J}^{\{n\rm AC\}} := \frac{\sin(2a)}{{\cal W}^{\{n \rm AC\}} +2 \tau  \partial_\tau {\cal W}^{\{n \rm AC\}}} \,.
\ee
As was the case for the AC vorticity, this expression is only valid near $a \simeq a_\star^\pm$, since the AC method requires the Domb--Sykes extrapolation results as input that are (currently) only trustworthy in that neighborhood. In contrast to the AC vorticity, however, the AC Jacobian~\eqref{eq:jacUV} is in general {\it non-zero} in the limit $a \to a_\star^+ = 0,\pm \pi$ for any $\tau <\tau_\star^{\smallL}(a_\star^+) = R_\star$, and thus, it can be used as a clean diagnostic to be evaluated {\it exactly at the blow-up location(s)}. In fact, for any $n>1$, the AC Jacobian is strictly positive for times before the blow-up, and is exactly zero at $\tau = R_\star$ at the locations~$a=a_\star^+$.

\begin{figure}
 \centering 
\includegraphics[width=0.6\textwidth]{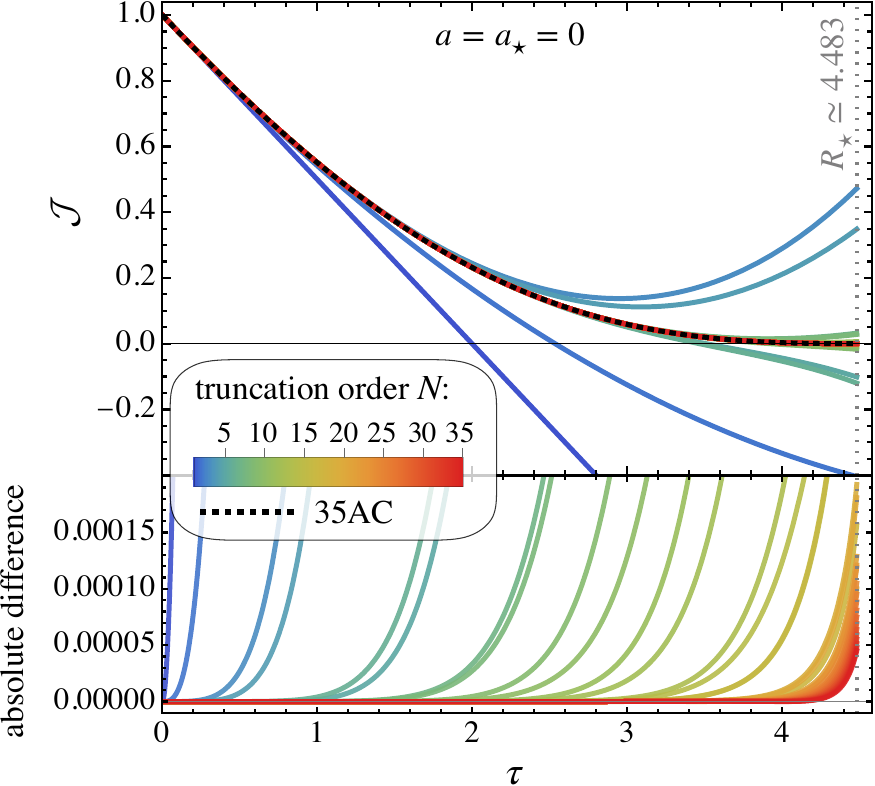}

\caption{Evolution of Jacobian ${\cal J}= \nabla_a {\cal Z}$ evaluated at the blow-up location $a=a_\star =0$. Shown are truncated series solutions based on the raw expansion (solid lines in various colors; based on Eq.\,\ref{eq:psi-sol}), as well as the AC prediction~\eqref{eq:jacUV} at truncation order $n=35$ (black dotted line). Curves are plotted for $0 \leq \tau \leq R_\star$. The bottom panel shows the absolute difference between ${\cal J}^{\{35\rm AC\}}$ and the truncated series solutions.
} \label{fig:UV-jac}
\end{figure}

In Fig.\,\ref{fig:UV-jac}, we show the temporal evolution of the Jacobian ${\cal J} = 1 + \nabla_a \mPsi$ evaluated at $a \to a_\star =0$, obtained with the AC model at truncation order $n=35$ (black dotted line), compared against raw series solutions at varying truncation orders (colors as indicated in the legend). The truncated series solutions converge towards the AC solution, with the expected trend that the highest  truncation orders retain accuracy for the longest times. It is, however, also noticeable that the convergence of the raw series expansion of the Jacobian is fairly slow, especially at times close to the blow-up. This slow convergence indicates that the HL model is (still) highly singular in nature, despite the usage of Lagrangian coordinates; for comparison, see e.g.\ \cite{Rampf:2017jan, Rampf:2021rqu, Rampf:2022apj} for different fluids where convergence of the Lagrangian series is significantly accelerated compared to its Eulerian one.

\section{Experiment: AC model and Beale--Kato--Majda criterion} \label{sec:BKM}

Here we employ the AC model derived in Sec.~\ref{sec:UV} and conduct a numerical experiment: We take the AC vorticity~\eqref{eq:WnUV} as a stand-in for the vorticity output of a Lagrangian grid-based numerical simulation, and assume that this surrogate output faithfully represents the true solution (within the validity domain of the AC construction). This surrogate field exhibits coordinate-dependent temporal singularities~$\tau_\star^{\smallL}(a) \in \mathbb{R}$ and associated exponents~$\nu^{\smallL}(a) \in \mathbb{R}$ (Fig.\,\ref{fig:DS-Rofa}). Our aim is to apply a standard global blow-up criterion---namely the Beale--Kato--Majda (BKM) condition~\cite{BKM-1984,1993CMaPh.155..277F}---to assess to which degree the nature of the underlying singularities can be recovered from these data.

For the 3D Euler case, the existence of a finite-time singularity can be decisively addressed using the BKM criterion, which states that the time integral of the $L^{\infty}$-norm of the vorticity diverges in finite time $t_s$, i.e., $\int_0^{t_s} \| \fett \omega \|_\infty \dd t= \infty$.
In contrast, for the HL model, it remains unclear whether every loss of regularity is a robust criterion for detection of finite-time blow-up \cite{Choi2017}. Nonetheless, as an intuitively motivated diagnostic, we may apply the following analogous blow-up condition to the HL model
\be \label{eq:BKMcon}
   \int_0^{{\cal T}_\star} \| {\cal W} \|_\infty \,\dd \tau = \infty \,,\qquad\qquad \qquad [\text{BKM~criterion~applied~to~HL~model}]
\ee
where ${\cal T}_\star$ is the estimate on the blow-up time, while ${\cal W}$ is the Lagrangian vorticity as defined through Eqs.\,\eqref{eqs:HL-Lag}.
The proposed experiment is successful if ${\cal T}_\star$ agrees with~$\tau_\star^{\smallL}(a=a_\star^+) = R_\star$. For this purpose, we replace ${\cal W}$ in~\eqref{eq:BKMcon} by the AC model ${\cal W}^{{\mathrm{15AC}}}$ (Eq.\,\eqref{eq:WnUV} for $n=15$), whose remainder term captures the non-analytic behavior of the HL model,
\be
  {\cal W}^{\rm asymp}(a, \tau)  \propto \left[ \tau_\star^{\smallL}(a) - \tau \right]^{\nuL(a)}  \,. 
\ee
We emphasize that the AC method only leads to meaningful predictions $\left| a - a_\star^\pm \right| \lesssim 0.4$ near the critical positions~$a_\star^\pm =  0\,, \pm \pi/2\,, \pm \pi$ (see Fig.\,\ref{fig:DS-Rofa}). However, since we are interested in the temporal evolution of the maxima of ${\cal W}^{\{\rm 15 AC\}}$ precisely in such neighborhoods, we restrict the spatial domain over which the norm is evaluated.

\begin{figure}
 \centering 
\includegraphics[width=0.65\textwidth]{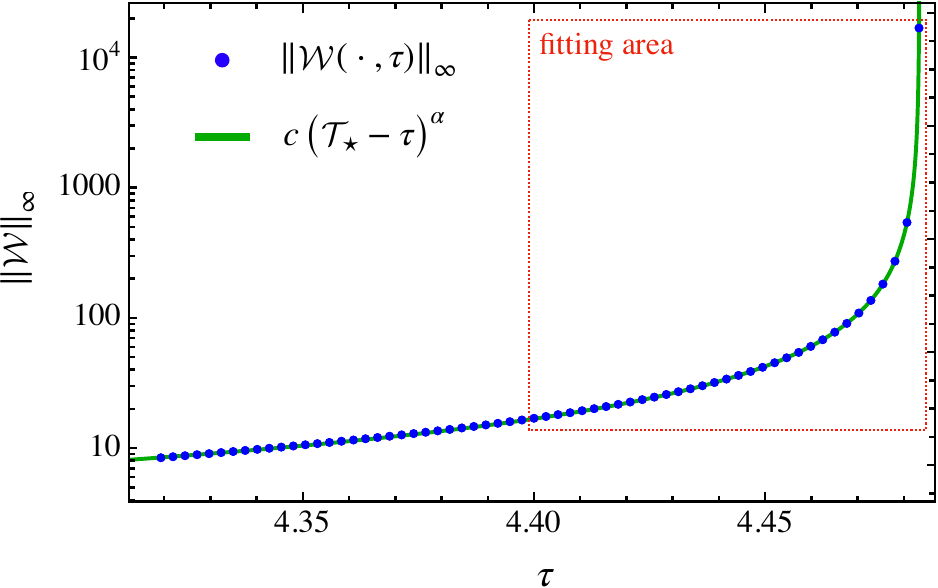}

\caption{Evolution of the maximum vorticity ${\cal W}$ (blue points), here obtained by sampling the AC model ${\cal W}^{\rm 15AC}$ on a Lagrangian mesh with 128 grid-points for the sub-domain $-0.1\leq a \leq 0.1$ around the blow-up location~$a=a_\star=0$. The power law-fit (green solid line; from BKM) recovers the blow-up time ${\cal T}_\star = R_\star \simeq 4.48339$ to six significant digits, but the measured exponent $\alpha \simeq -1.01$ clearly deviates from the local singularity exponent~$\nu_\star \simeq -1.51$. See Tab.\,\ref{tab:conv} for a summary of results at varying spatial resolution.
} \label{fig:BKM}
\end{figure}

Since we treat the AC prediction as a surrogate for a direct numerical simulation (DNS), we also introduce an artificial spatial discretization to mimic the resolution constraints inherent to a DNS (noting that this probes only collocation errors, as the AC model is assumed to be free from subgrid limitations). For this, we sample ${\cal W}^{{\mathrm{15AC}}}$ over the sub-domain $-0.1 \leq a \leq 0.1$ using a uniformly spaced grid with a variable number of grid points~$N_z= 32,64,128,512,1024,\infty$, where the last entry denotes the continuous case. In Fig.\,\ref{fig:BKM} we show the results of the maximum vorticity as a function of time (blue points), for $N_z= 128$ (results for higher resolutions look indistinguishable). As expected---and in line with previous observations (Fig.\,\ref{fig:UV-vorticity})---the maximal vorticity exhibits signs of a blow-up. To quantify this, we apply the non-analytic model 
\be \label{eq:Wnorm-model}
  \| {\cal W} \|_{\infty, \rm model} = c \left( {\cal T}_\star - \tau \right)^\alpha
\ee
and determine the unknown parameters through a non-linear least-squares fit, using 32 temporal data points uniformly distributed between $4.4000 \leq \tau \leq 4.4833$ (red box in Fig.\,\ref{fig:BKM}), immediately preceding the blow-up at $\tau =R_\star \simeq 4.48339$.

It is important to note (again) that, in the AC model, the temporal singularity $\tau_\star^{\smallL}(a)$ varies with position and is approximately parabolic near $a=0$ (Fig.\,\ref{fig:DS-Rofa}). For the last temporal data point used in the fit, at $\tau_{\rm max} = 4.4833$, the maximum of the vorticity occurs at $a_{\tau_{\rm max}} \simeq 0.002754$ 
for~$N_z=\infty$, yielding a corresponding local singularity time $\tau_\star^{\smallL}(a_{\tau_{\rm max}}) \simeq 4.48343$, which is slightly larger than the true blow-up time $R_\star$. This is a subtle but important point: although the AC model serves as a perfect surrogate, the vorticity maximum can only be tracked up until time $\tau_{\rm max}<R_\star$ when it is confined to a finite distance away from the point of singularity ($a=0$). 
Consequently, the BKM fit must recover the blow-up time by extrapolation rather than by direct sampling of the singular regime.
Furthermore, this problem can be aggravated due to the resolution constraints imposed by finite $N_z$.

\begin{table}[t]
\centering
\begin{tabular}{c|ccccccc}
\hline\hline
$N_z$ 
  & $32$ & $64$ & $128$ & $256$ & $512$ & $1024$ & $\infty$ \\ 
\hline
$\alpha$ 
  & $-1.00879$\phantom{+} & $-1.01128$\phantom{+} & $-1.01112$\phantom{+} & $-1.01136$\phantom{+} 
  & $-1.01127$\phantom{+} & $-1.01129$\phantom{+} & $-1.01129$ \\
${\cal T}_\star$ 
  & $\phantom{+}4.83440$\phantom{+} & $\phantom{+}4.48339$\phantom{+} & $\phantom{+}4.48339$\phantom{+} & $\phantom{+}4.48339$\phantom{+}
  & $\phantom{+}4.48339$\phantom{+} & $\phantom{+}4.48339$\phantom{+} & $\phantom{+}4.48339$ \\
\hline\hline
\end{tabular}
\caption{Convergence of the fitted singularity exponent $\alpha$ and blow-up time ${\cal T}_\star$ based on the model~\eqref{eq:Wnorm-model} with increasing grid resolution~$N_z$ for the sub-domain $-0.1\leq a \leq 0.1$, centered about the blow-up location~$a_\star=0$.}
\label{tab:conv}
\end{table}

In Fig.\,\ref{fig:BKM}, we show the resulting fit (green solid line) with best fit values obtained for $N_z = 128$; the corresponding fits for all resolutions are summarized in Table\,\ref{tab:conv}. For $N_z \geq 64$, the extrapolation reveals ${\cal T}_\star = R_\star$ accurate to 6 significant digits, thereby indicating that, indeed, the blow-up time may be retrieved from sufficiently resolved DNS data. In contrast, the behavior for the extracted singularity exponent differs: We observe a slower convergence and reach the same 6-digit accuracy only for $N_z \geq 1024$. More importantly, the retrieved exponent $\alpha\simeq  -1.01129$ is markedly different from the input value of our surrogate DNS where $\nu_\star \simeq -1.51487$ corresponding to $a=0$. This mismatch can be explained as follows. The BKM-based power-law fit only probes the supremum norm which is insensitive to the underlying spatial structure of the singularity. Specifically, in the present case, the position of the maximum drifts towards the origin as $\tau \to {\cal T}_\star$, so that the extracted exponent~$\alpha$ represents an effective exponent arising from a succession of spatial locations that temporarily dominate the maximum. While the underlying mechanism seems evident, at this stage we are unable to derive the value of the measured effective exponent from first principles.

\section{Lagrangian singularity theory}\label{sec:sing-theory}

We have seen that, within an Eulerian formulation, the MR extrapolation method draws the picture of a singular landscape in complex time (Fig.\,\ref{fig:sing}). Ultimately however, methods such as MR are phenomenological in nature where it is difficult to quantify extrapolation errors. Here, we show that the landscape of complex-time singularities, obtained above with the Eulerian formulation, can be derived by a Lagrangian theory. In the present realization, this theory relies on a Taylor-series representation whose radius of convergence spans the complex-time disc of radius~$R_\star \simeq 4.483$, unlike Eulerian series solutions which are constrained by the complex singularity at $R_{\rm inf} \simeq 1.588$. Thus, predictions from the Lagrangian theory are only limited by truncation errors of the convergent series, which, as we show, are well-controlled.

We first describe the construction of the theory in the ideal case in which the mathematically exact solution for the characteristic map ${\cal Z}(a,\tau)$ is available, and subsequently address truncation errors. In the idealized setting, the construction  follows closely the theory for the 1D inviscid Burgers equation~\cite{Rampf:2022apj}, for which truncation errors are absent. Therefore, the incorporation of truncation constitutes a new and essential ingredient of the present theory.

The central idea underlying the theory is that finite-time singularities in the HL model are triggered when multiple fluid particles approach the same Eulerian position~$z$. In Lagrangian coordinates, the last statement is equivalent to searching for the vanishing of the Jacobian of the Lagrangian map~${\cal J}= \partial {\cal Z}/\partial a$, which occurs first at time $\tau = R_\star \simeq 4.48339$ at the Lagrangian locations~$a=a_\star^+ = 0,\pm \pi$ (see Sec.\,\ref{sec:asyLag} for details). Allowing the time variable to take complex values
\be \label{eq:complextau}
   \fett \tau := |\fett \tau| \,{\rm e}^{\ii \Theta} 
\ee
instead, the condition of a vanishing Jacobian can also be satisfied at complex times with a modulus of $\fett \tau$ that is smaller than $R_\star$. Indeed, by formally complexifying ${\cal J}$, ${\cal Z}$ and $a$, denoted respectively by $\fett {\cal J}$, $\fett {\cal Z}$ and $\fett a$, the root-finding task with complex solutions~$\fett a_i$ takes the form
\be \label{eq:im-roots}
   \fett a = \fett a_i \,: \qquad \fett {\cal J}(\fett a, \fett \tau) := \frac{\partial \fett {\cal Z}}{\partial \fett a} \stackrel ! = \fett 0\,, \qquad |\fett \tau| < R_\star \,.
\ee
For the present HL model and chosen initial data, a numerical root search applied to the Taylor series of the Jacobian reveals that $i=4$, i.e., we have four such roots. To manage the complexity of the root search, we adopt a parametric approach where we solve for the $\fett a_i$ as a function of $|\fett \tau|$ for varying phase $\Theta$. Additionally, the (complex) time evolution of these roots depend in general also on the truncation order employed in the Jacobian. However, as we show below, the truncation errors in the root search are vanishingly small for phases $\Theta \simeq \pm \pi/2$, namely of the order $10^{-8}$ for truncation orders~$N \gtrsim 33$ (see Fig.\,\ref{fig:theory-trunc} below). Hence, to demonstrate the mathematical significance of the roots, we first discuss the special case $\Theta = \pi/2$ (while setting $N=36$), and generalize our method to arbitrary $\Theta$ later on, including a detailed discussion on truncation errors.

\begin{figure}
 \centering 
\includegraphics[width=0.97\textwidth]{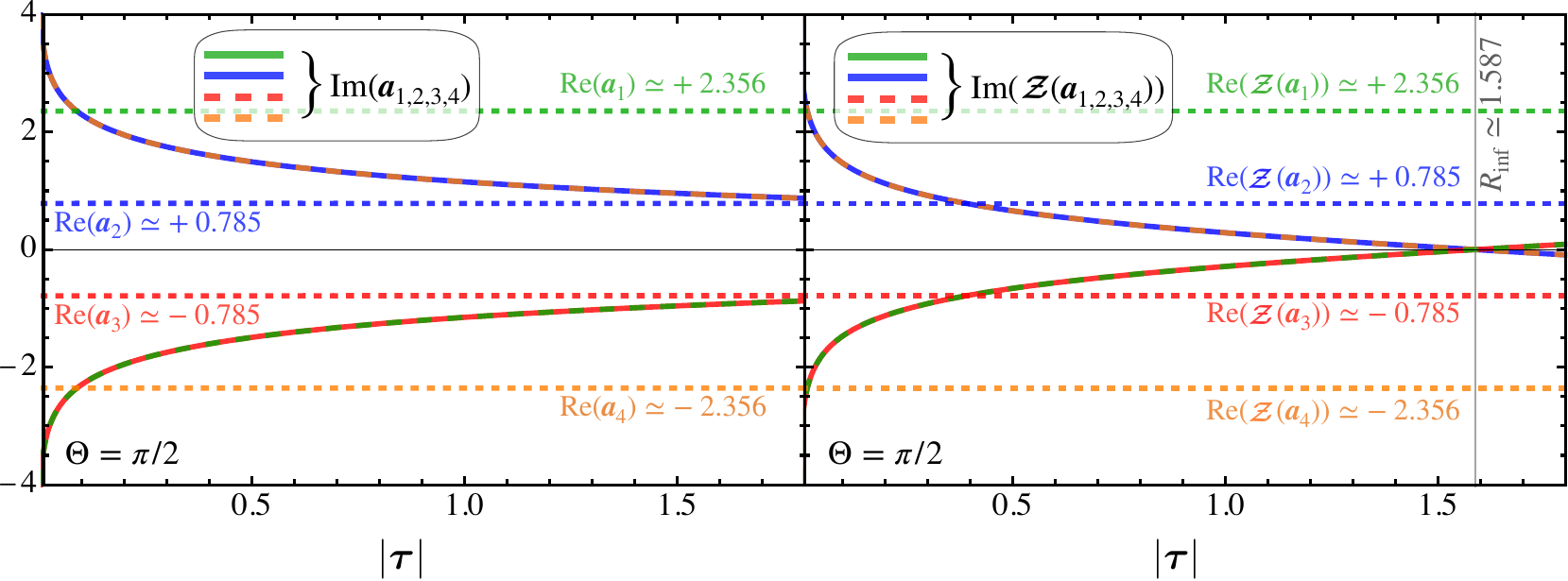}

\caption{{\it Left panel:}  Evolution of the four roots~$\fett a_{1,2,3,4}$ of the complexified Jacobian~$\fett {\cal J}$ as a function of~$\fett \tau = |\fett \tau|\exp(\ii \Theta)$ for the phase choice $\Theta = \pi/2$.
{\it Right panel:} Corresponding evolution of the complexified position~$\fett{\cal Z}$ evaluated at these roots. According to the theory, the first instance of ${\rm Im}\!\left[\fett{\cal Z}(\fett a_i)\right] = 0$ signals a singularity at $\fett \tau = \fett \tau_{\star,i}$ as seen within the Eulerian formulation. For the present choice of phase, the singularities closest to the origin have a modulus of $|\fett \tau| = R_{\rm inf} \simeq 1.587$ and manifest first at the Eulerian positions ${\rm Re}\!\left[\fett{\cal Z}(\fett a_{1,2,3,4})\right] \simeq \pm 0.785, \pm2.356$ (cf.\ left panel of Fig.\,\ref{fig:sing}). The root finding uses the $\fett \tau$-time Taylor series of the characteristic map at truncation order~$N=36$ as input.
} \label{fig:aroots-Z}
\end{figure}
%

In the left panel of Fig.\,\ref{fig:aroots-Z}, we show the four roots as a function of the modulus of complex time~$|\fett \tau|$ for $\Theta = \pi/2$. Two of the four roots share the same imaginary part that varies in time. Their real parts remain constant in this particular setting, which is not a generic feature but does not entail any additional complications.

Having this observation in mind, we now exploit the key idea of the theory, namely that these roots trigger non-analyticity, as observed at the real Eulerian coordinates, when the imaginary part of the complex position evaluated at a given root vanishes for the first time~\cite{Rampf:2022apj}; i.e., we impose 
\be \label{eq:realness}
   \fett \tau = \fett \tau_{\star,i} \,: \qquad {\rm Im}\left( \fett {\cal Z}(\fett a = \fett a_i, \fett \tau) \right) \stackrel ! = 0 \,.
\ee
Then, in a final step---not required for the present setup---we set $\fett \tau_\star = {\rm inf} \{ \fett \tau_{\star,i} \}$, thereby selecting the singularity closest to the origin of time as being the physically relevant one (again, unnecessary in the present case as $\fett \tau_\star = \fett \tau_{\star,i}$ for $i=1,2,3,4$; see discussion below).

The above methodology is visualized in the right panel of Fig.\,\ref{fig:aroots-Z}, where we fix the phase of~$\fett \tau$ to  $\Theta = \pi/2$ as before. At early times~$|\fett \tau|$, the imaginary parts of $\fett {\cal Z}$ at the root positions $\fett a_{1,2,3,4}$ are non-zero and (for the present setting) cross zero at the same instant $|\fett \tau| = R_{\rm inf} \simeq 1.587$. This is precisely the minimal radius of convergence observed at the Eulerian positions $z = \pm \pi/4, \pm 3\pi/4 \simeq  \pm 0.785, \pm2.356$ (cf.\ left panel in Fig.\,\ref{fig:sing}), which lead to the observation of the early-time tygers~(Fig.\,\ref{fig:fluid-variables}). Moreover, the theory not only predicts the locations of the singularities, but also identifies where these singularities first become relevant in Eulerian space: This information is obtained by evaluating the real parts ${\rm Re}\!\left[\fett{\cal Z}(\fett a_{1,2,3,4})\right]$, which---up to tiny numerical root-finding uncertainties---yield exactly the same Eulerian positions stated above.

The above considerations apply to any choice of phase~$\Theta$ of $\fett \tau$, thereby allowing us to reconstruct the singular structure of the Eulerian formulation within the temporal domain (modulo truncation errors, see below). For this, we vary $\Theta$ and search for the corresponding $|\fett \tau_\star|$ that satisfies the realness condition~\eqref{eq:realness}, thereby leading to the results shown in the right panel of Fig.\,\ref{fig:sing}. Similarly, by drawing respectively $|\fett \tau_\star|$  and $\Theta$ against ${\rm Re}\left( \fett {\cal Z}(\fett a_i, \fett \tau_\star) \right)$, one retrieves the theoretical predictions shown in the left panel of Fig.\,\ref{fig:sing}.

Finally, we discuss truncation errors in the present setup. Both the Lagrangian root finding~\eqref{eq:im-roots} as well as the evaluation of the realness condition~\eqref{eq:realness} require the truncated series ${\fett {\cal Z}}^{\{N\}}(\fett a,\fett \tau) := \fett a+ \sum_{n=1}^N \fett{\mPsi}_n(\fett a) \,\fett \tau^n$ as input, where $\fett{\mPsi}_n$ are the formally complexified series coefficients of~\eqref{eq:psi}. 
\begin{figure}
 \centering 
\includegraphics[width=0.93\textwidth]{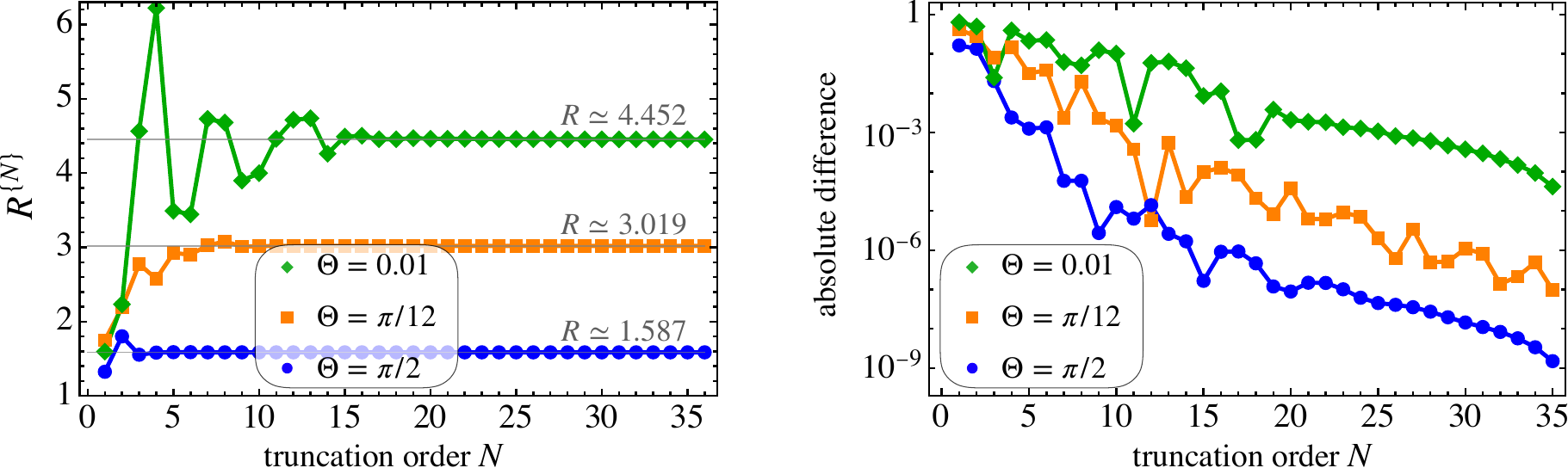}

\caption{Truncation errors when solving~Eq.\,\eqref{eq:realness} approximately with ${\fett {\cal Z}}^{\{N\}}(\fett a,\fett \tau) = \fett a+ \sum_{n=1}^N \fett{\mPsi}_n(\fett a) \,\fett \tau^n$.
{\it Left panel:}  Predicted modulus $R= R^{\{N\}}$ of the singularity $\fett \tau_\star = R \exp(\ii \Theta)$ as a function of truncation order $N$, shown for three phase choices  $\Theta = 0.01, \pi/12, \pi/2$ (respectively shown with green diamonds, orange squares and blue circles). 
{\it Right panel:} Absolute relative difference with respect to~$R^{\{36\}}$ at different truncation orders for the three phase choices shown in the legend. 
} \label{fig:theory-trunc}
\end{figure}
In the left panel of Fig.\,\ref{fig:theory-trunc} we show the predicted modulus~$R$ of~$\fett \tau_\star = R \exp(\ii \Theta)$ as a function of truncation order, for three choices of the phase~$\Theta = 0.01,\pi/12,\pi/2$. The results for $\Theta = \pi/2$ (blue circles) correspond to singularities located precisely on the imaginary time axis (${\rm Re}(\fett \tau_\star)=0$); in this case, the solution rapidly converges to a stable solution, with truncation errors being below $10^{-4}$, for $N \gtrsim 7$, relative to our benchmark solution (right panel in Fig.\,\ref{fig:theory-trunc}; note that the MR method utilized Taylor coefficients up to order~36 to determine $R_{\rm inf}$ with four-digit accuracy). The fast convergence of the Lagrangian theory arises because the corresponding singularity, $R=R_{\rm inf}\simeq 1.587$, lies well within the radius of convergence of the Lagrangian series, which we have estimated with $R=R_\star\simeq 4.483$ (Sec.\,\ref{sec:asyLag}); thus, as generally expected for convergent series representations, and as observed in Fig.\,\ref{fig:theory-trunc}, truncation errors become quickly negligible.
 
Truncation errors become more relevant when the predicted modulus of $\fett \tau_\star$ approaches the boundary of convergence. This is shown in Fig.\,\ref{fig:theory-trunc} for $\Theta = \pi/12$ (orange squares) and in particular for $\Theta =0.01$ (green diamonds) where truncation errors decay considerably slower than in the previously discussed cases. Nevertheless, even for such singularities which are located close to the real time axis, theoretical results are accurate to about four digits for truncation orders $N \gtrsim 35$ and, crucially, the prediction is obtained without any extrapolation.

\section{Conclusions}\label{sec:concl}

Detecting finite-time blow-up by direct numerical simulations is inherently challenging. Numerical precision deteriorates near singular points, and one is therefore forced to extrapolate from data obtained sufficiently close to---but still clearly off---the blow-up. The situation changes significantly if one knows a `desingularization' transformation for the problem at hand which, roughly speaking, turns infinities into zeros, thereby making the singularity amenable to direct evaluation.

In this paper, we have investigated the singular structure of the one-dimensional Hou--Luo (HL) model and shown that the introduction of Lagrangian (characteristic) coordinates provides precisely such a desingularization. More specifically, the characteristic mapping from the initial to the current fluid particle position along the axial direction acts as a desingularization transformation, whose corresponding Jacobian vanishes at the blow-up location (Fig.\,\ref{fig:UV-jac}). This Jacobian measures the axial separation of fluid particles and collapses when multiple particles converge to the same Eulerian position, signaling the loss of monotonicity of the mapping and an impending bifurcation.

In Lagrangian coordinates, the HL blow-up persists, but the structure of the closest singularities is considerably simplified: Instead of the continuous eye-shaped proliferation of singularities present in Eulerian coordinates (Fig.\,\ref{fig:sing}), only two discrete singularities remain relevant in the Lagrangian formulation, corresponding to the real blow-up time in the positive and negative temporal directions. This simplified singular structure comes with several important consequences that we exploit in the present paper. For instance, the Lagrangian time-Taylor series solutions~\eqref{eq:psi-sol}--\eqref{eq:W} remain mathematically meaningful until the blow-up time, whereas the Eulerian counterparts~\eqref{eqs:eul_ttc_exp} lose regularity at earlier times, triggered by complex-time singularities (Fig.\,\ref{fig:fluid-variables}). These results are substantiated by asymptotic analyses in both Eulerian and Lagrangian coordinates, based on the first 72 $t$-time Taylor coefficients, which we have evaluated symbolically (see App.\,\ref{app:MR} for further asymptotic results and App.\,\ref{app:recs} for  accompanying codes).

Furthermore, we have developed a simple model for the Lagrangian vorticity~${\cal W}$ that resolves the spatio-temporal regime near the blow-up to good accuracy (Fig.\,\ref{fig:UV-vorticity}). This model is obtained through {\it asymptotic completion} (AC), in which the remainder of the truncated series for~${\cal W}$ encapsulates the asymptotic singularities of the solution (Eq.\,\ref{eq:WnUV}). We exploit this AC model as a testing ground to assess the extent to which the local singularity information can be retrieved by applying the global Beale--Kato--Majda (BKM) criterion to the AC vorticity. By fitting a space-independent non-analytic model to the time-dependence of the $L^\infty$ norm $\|{\cal W}\|_\infty$, until shortly before the blow-up, we find that the extrapolation recovers the blow-up time with high precision, but washes out the local singularity exponent as the spatial supremum is insensitive to the local structure of the singularity (Fig.\,\ref{fig:BKM} and Table\,\ref{tab:conv}).

The vanishing of the Jacobian serves as a clean diagnostic for detecting the HL blow-up. The Lagrangian singularity theory developed here formally complexifies the root-finding problem for the Jacobian in both space and time, and predicts which complex-time singularities manifest at a given real-valued Eulerian position~$z$ (Figs.\,\ref{fig:sing} and~\ref{fig:aroots-Z}). Conceptually, the present theory constitutes a generalization of the one recently developed for the 1D inviscid Burgers \cite{Rampf:2022apj}, in that it now accommodates series solutions in the root finding procedure (Fig.\,\ref{fig:theory-trunc}), rather than relying on the exact closed-form solution available in the Burgers case. Accounting for the resulting rounding errors represents an important extension, and renders the singularity theory applicable to more complex hydrodynamical equations.

We emphasize that it is the Eulerian formulation alone that exhibits (early-time or tyger) resonances, as previously analyzed in a range of fluids  \cite{Kolluru2022PhRvE,KolluruPandit2024,Podvigina2016,Rampf:2022apj}. This observation suggests that numerical schemes for such fluids may benefit from Lagrangian implementations as well. While much recent work in fluid dynamics has focused on Eulerian approaches, Lagrangian descriptions offer complementary and often crucial insights in a range of settings, including cosmology and plasma physics~\cite{Rampf:2021rqu}. Lagrangian formulations for the 3D axisymmetric Euler equations and beyond do exist (e.g.\ \cite{2008PhyD..237.1951M,Zheligovsky:2013eca,Podvigina2016,Besse2017,Hertel2022}), but we believe their implications for singularity detection have not yet been explored systematically. The present work therefore highlights the added value of combining Eulerian and Lagrangian descriptions in obtaining robust evidence for or against singularity formation in hydrodynamical flows.

In the present paper, the developed theory employs single-step series solutions to solve for the complex roots. However, the solution strategy could be adapted given a specific problem at hand, e.g., either by using a multi-time stepping algorithm such as the Cauchy--Lagrange method \cite{Podvigina2016,Hertel2022}, or by developing a direct numerical simulation technique in complexified space. Moreover, by now the characteristic eye-shaped structure of singularities has been observed in two independent fluid models, which indicates that this shape could be somewhat generic. More work is required to assess this possibility, which we speculate could trigger the development of new singularity detection methods.

Our theory may be straightforwardly applied to other fluids, in particular incompressible flows, where the Jacobian of the direct Lagrangian mapping is constrained to unity throughout evolution. In this context, recall that the HL model is compressible, although its parent system, the 3D axisymmetric incompressible Euler (3DAE) equations, is indeed incompressible. The apparent conundrum is resolved by the fact, that the Jacobian appearing in the HL formulation is not associated with such a volume-preserving map, but arises from a reduced characteristic description (3DAE evaluated at the wall). Consequently, when applying the present theory to incompressible flows, it is not the determinant of the Jacobian of the volume-preserving map that vanishes, but rather specific matrix element(s) that may degenerate, without violating incompressibility. This distinction clarifies the role of the Jacobian in the theory and shows that the underlying mechanism for singularity formation may extend naturally beyond compressible flows.

While the HL model is one-dimensional, our theory can be straightforwardly applied to fluids in more space dimensions. This would require the development, therein, of a Lagrangian-coordinates formulation (if not already present in the literature), followed by an asymptotic analysis in order to probe the validity regime of the Lagrangian solutions. In higher dimensions, the complex root-finding problem underlying the singularity theory becomes intrinsically higher-dimensional and therefore more involved; however, this does not constitute a fundamental limitation of the approach. In practice, the analysis is controlled by the singularity of minimal modulus, and only this root needs to be identified in order to detect the physically relevant singularity.

As a next but important step, we will apply the present singularity framework to 2D Boussinesq flows \cite{ElgindiJeong2020,Chen-Hou-2021CMaPh.383.1559C} and to the 3D inviscid Taylor--Green vortex \cite{Morf1980,brachet,Brachet1992PhFlA...4.2845B}. These systems provide natural testbeds for assessing the robustness and generality of the developed methods, and, in the longer term, for contributing to systematic progress on the search for finite-time singularities in hydrodynamical equations.

\begin{acknowledgments}
We thank Uriel Frisch and Rahul Pandit for related discussions, and Samriddhi Sankar Ray for useful comments on the manuscript.
\end{acknowledgments}

\bibliographystyle{apsrev4-2} 
\bibliography{references}

\begin{thebibliography}{51}%
\makeatletter
\providecommand \@ifxundefined [1]{%
 \@ifx{#1\undefined}
}%
\providecommand \@ifnum [1]{%
 \ifnum #1\expandafter \@firstoftwo
 \else \expandafter \@secondoftwo
 \fi
}%
\providecommand \@ifx [1]{%
 \ifx #1\expandafter \@firstoftwo
 \else \expandafter \@secondoftwo
 \fi
}%
\providecommand \natexlab [1]{#1}%
\providecommand \enquote  [1]{``#1''}%
\providecommand \bibnamefont  [1]{#1}%
\providecommand \bibfnamefont [1]{#1}%
\providecommand \citenamefont [1]{#1}%
\providecommand \href@noop [0]{\@secondoftwo}%
\providecommand \href [0]{\begingroup \@sanitize@url \@href}%
\providecommand \@href[1]{\@@startlink{#1}\@@href}%
\providecommand \@@href[1]{\endgroup#1\@@endlink}%
\providecommand \@sanitize@url [0]{\catcode `\\12\catcode `\$12\catcode
  `\&12\catcode `\#12\catcode `\^12\catcode `\_12\catcode `\%12\relax}%
\providecommand \@@startlink[1]{}%
\providecommand \@@endlink[0]{}%
\providecommand \url  [0]{\begingroup\@sanitize@url \@url }%
\providecommand \@url [1]{\endgroup\@href {#1}{\urlprefix }}%
\providecommand \urlprefix  [0]{URL }%
\providecommand \Eprint [0]{\href }%
\providecommand \doibase [0]{https://doi.org/}%
\providecommand \selectlanguage [0]{\@gobble}%
\providecommand \bibinfo  [0]{\@secondoftwo}%
\providecommand \bibfield  [0]{\@secondoftwo}%
\providecommand \translation [1]{[#1]}%
\providecommand \BibitemOpen [0]{}%
\providecommand \bibitemStop [0]{}%
\providecommand \bibitemNoStop [0]{.\EOS\space}%
\providecommand \EOS [0]{\spacefactor3000\relax}%
\providecommand \BibitemShut  [1]{\csname bibitem#1\endcsname}%
\let\auto@bib@innerbib\@empty
\bibitem [{\citenamefont {{Eyink}}(2024)}]{Eyink-onsager}%
  \BibitemOpen
  \bibfield  {author} {\bibinfo {author} {\bibfnamefont {G.}~\bibnamefont
  {{Eyink}}},\ }\href {https://doi.org/10.1017/jfm.2024.415} {\bibfield
  {journal} {\bibinfo  {journal} {J. Fluid Mech.}\ }\textbf {\bibinfo {volume}
  {988}},\ \bibinfo {eid} {P1} (\bibinfo {year} {2024})},\ \Eprint
  {https://arxiv.org/abs/2404.10084} {arXiv:2404.10084} \BibitemShut {NoStop}%
\bibitem [{\citenamefont {Luo}\ and\ \citenamefont {Hou}(2014)}]{HouLuo2014}%
  \BibitemOpen
  \bibfield  {author} {\bibinfo {author} {\bibfnamefont {G.}~\bibnamefont
  {Luo}}\ and\ \bibinfo {author} {\bibfnamefont {T.~Y.}\ \bibnamefont {Hou}},\
  }\href {https://doi.org/10.1073/pnas.1405238111} {\bibfield  {journal}
  {\bibinfo  {journal} {Proc. Natl. Acad. Sci. U.S.A.}\ }\textbf {\bibinfo
  {volume} {111}},\ \bibinfo {pages} {12968} (\bibinfo {year} {2014})},\
  \Eprint {https://arxiv.org/abs/1310.0497} {arXiv:1310.0497} \BibitemShut
  {NoStop}%
\bibitem [{\citenamefont {{Beale}}\ \emph {et~al.}(1984)\citenamefont
  {{Beale}}, \citenamefont {{Kato}},\ and\ \citenamefont {{Majda}}}]{BKM-1984}%
  \BibitemOpen
  \bibfield  {author} {\bibinfo {author} {\bibfnamefont {J.~T.}\ \bibnamefont
  {{Beale}}}, \bibinfo {author} {\bibfnamefont {T.}~\bibnamefont {{Kato}}},\
  and\ \bibinfo {author} {\bibfnamefont {A.}~\bibnamefont {{Majda}}},\ }\href
  {https://doi.org/10.1007/BF01212349} {\bibfield  {journal} {\bibinfo
  {journal} {Commun. Math. Phys.}\ }\textbf {\bibinfo {volume} {94}},\ \bibinfo
  {pages} {61} (\bibinfo {year} {1984})}\BibitemShut {NoStop}%
\bibitem [{\citenamefont {{Deng}}\ \emph {et~al.}(2005)\citenamefont {{Deng}},
  \citenamefont {{Hou}},\ and\ \citenamefont
  {{Yu}}}]{Deng-Hou-Yu-2004math.ph...2032D}%
  \BibitemOpen
  \bibfield  {author} {\bibinfo {author} {\bibfnamefont {J.}~\bibnamefont
  {{Deng}}}, \bibinfo {author} {\bibfnamefont {T.~Y.}\ \bibnamefont {{Hou}}},\
  and\ \bibinfo {author} {\bibfnamefont {X.}~\bibnamefont {{Yu}}},\ }\href
  {https://doi.org/10.1081/PDE-200044488} {\bibfield  {journal} {\bibinfo
  {journal} {Commun. Partial Differ. Equ.}\ }\textbf {\bibinfo {volume} {30}},\
  \bibinfo {pages} {225} (\bibinfo {year} {2005})},\ \Eprint
  {https://arxiv.org/abs/math-ph/0402032} {arXiv:math-ph/0402032} \BibitemShut
  {NoStop}%
\bibitem [{\citenamefont {{Hou}}\ and\ \citenamefont
  {{Li}}(2006)}]{HouRuo2006JNS....16..639H}%
  \BibitemOpen
  \bibfield  {author} {\bibinfo {author} {\bibfnamefont {T.~Y.}\ \bibnamefont
  {{Hou}}}\ and\ \bibinfo {author} {\bibfnamefont {R.}~\bibnamefont {{Li}}},\
  }\href {https://doi.org/10.1007/s00332-006-0800-3} {\bibfield  {journal}
  {\bibinfo  {journal} {J. Nonlinear Sci.}\ }\textbf {\bibinfo {volume} {16}},\
  \bibinfo {pages} {639} (\bibinfo {year} {2006})},\ \Eprint
  {https://arxiv.org/abs/math-ph/0602051} {arXiv:math-ph/0602051} \BibitemShut
  {NoStop}%
\bibitem [{\citenamefont {Barkley}(2020)}]{Barkley2020}%
  \BibitemOpen
  \bibfield  {author} {\bibinfo {author} {\bibfnamefont {D.}~\bibnamefont
  {Barkley}},\ }\href {https://doi.org/10.1098/rspa.2020.0348} {\bibfield
  {journal} {\bibinfo  {journal} {Proc. R. Soc. Lond. A}\ }\textbf {\bibinfo
  {volume} {476}},\ \bibinfo {pages} {20200348} (\bibinfo {year} {2020})},\
  \Eprint {https://arxiv.org/abs/1902.05993} {arXiv:1902.05993} \BibitemShut
  {NoStop}%
\bibitem [{\citenamefont {{Fehn}}\ \emph {et~al.}(2022)\citenamefont {{Fehn}},
  \citenamefont {{Kronbichler}}, \citenamefont {{Munch}},\ and\ \citenamefont
  {{Wall}}}]{Fehn2022JFM...932A..40F}%
  \BibitemOpen
  \bibfield  {author} {\bibinfo {author} {\bibfnamefont {N.}~\bibnamefont
  {{Fehn}}}, \bibinfo {author} {\bibfnamefont {M.}~\bibnamefont
  {{Kronbichler}}}, \bibinfo {author} {\bibfnamefont {P.}~\bibnamefont
  {{Munch}}},\ and\ \bibinfo {author} {\bibfnamefont {W.~A.}\ \bibnamefont
  {{Wall}}},\ }\href {https://doi.org/10.1017/jfm.2021.1003} {\bibfield
  {journal} {\bibinfo  {journal} {J. Fluid Mech.}\ }\textbf {\bibinfo {volume}
  {932}},\ \bibinfo {eid} {A40} (\bibinfo {year} {2022})}\BibitemShut {NoStop}%
\bibitem [{\citenamefont {{Sulem}}\ \emph {et~al.}(1983)\citenamefont
  {{Sulem}}, \citenamefont {{Sulem}},\ and\ \citenamefont {{Frisch}}}]{sulem}%
  \BibitemOpen
  \bibfield  {author} {\bibinfo {author} {\bibfnamefont {C.}~\bibnamefont
  {{Sulem}}}, \bibinfo {author} {\bibfnamefont {P.-L.}\ \bibnamefont
  {{Sulem}}},\ and\ \bibinfo {author} {\bibfnamefont {H.}~\bibnamefont
  {{Frisch}}},\ }\href {https://doi.org/10.1016/0021-9991(83)90045-1}
  {\bibfield  {journal} {\bibinfo  {journal} {J. Comput. Phys.}\ }\textbf
  {\bibinfo {volume} {50}},\ \bibinfo {pages} {138} (\bibinfo {year}
  {1983})}\BibitemShut {NoStop}%
\bibitem [{\citenamefont {{Brachet}}\ \emph
  {et~al.}(1983{\natexlab{a}})\citenamefont {{Brachet}}, \citenamefont
  {{Meiron}}, \citenamefont {{Orszag}}, \citenamefont {{Nickel}}, \citenamefont
  {{Morf}},\ and\ \citenamefont {{Frisch}}}]{brachet}%
  \BibitemOpen
  \bibfield  {author} {\bibinfo {author} {\bibfnamefont {M.~E.}\ \bibnamefont
  {{Brachet}}}, \bibinfo {author} {\bibfnamefont {D.~I.}\ \bibnamefont
  {{Meiron}}}, \bibinfo {author} {\bibfnamefont {S.~A.}\ \bibnamefont
  {{Orszag}}}, \bibinfo {author} {\bibfnamefont {B.~G.}\ \bibnamefont
  {{Nickel}}}, \bibinfo {author} {\bibfnamefont {R.~H.}\ \bibnamefont
  {{Morf}}},\ and\ \bibinfo {author} {\bibfnamefont {U.}~\bibnamefont
  {{Frisch}}},\ }\href {https://doi.org/10.1017/S0022112083001159} {\bibfield
  {journal} {\bibinfo  {journal} {J. Fluid Mech.}\ }\textbf {\bibinfo {volume}
  {130}},\ \bibinfo {pages} {411} (\bibinfo {year}
  {1983}{\natexlab{a}})}\BibitemShut {NoStop}%
\bibitem [{\citenamefont {{Brachet}}\ \emph {et~al.}(1992)\citenamefont
  {{Brachet}}, \citenamefont {{Meneguzzi}}, \citenamefont {{Vincent}},
  \citenamefont {{Politano}},\ and\ \citenamefont
  {{Sulem}}}]{Brachet1992PhFlA...4.2845B}%
  \BibitemOpen
  \bibfield  {author} {\bibinfo {author} {\bibfnamefont {M.~E.}\ \bibnamefont
  {{Brachet}}}, \bibinfo {author} {\bibfnamefont {M.}~\bibnamefont
  {{Meneguzzi}}}, \bibinfo {author} {\bibfnamefont {A.}~\bibnamefont
  {{Vincent}}}, \bibinfo {author} {\bibfnamefont {H.}~\bibnamefont
  {{Politano}}},\ and\ \bibinfo {author} {\bibfnamefont {P.~L.}\ \bibnamefont
  {{Sulem}}},\ }\href {https://doi.org/10.1063/1.858513} {\bibfield  {journal}
  {\bibinfo  {journal} {Phys. Fluids A}\ }\textbf {\bibinfo {volume} {4}},\
  \bibinfo {pages} {2845} (\bibinfo {year} {1992})}\BibitemShut {NoStop}%
\bibitem [{\citenamefont {{Cichowlas}}\ and\ \citenamefont
  {{Brachet}}(2005)}]{cichowlas}%
  \BibitemOpen
  \bibfield  {author} {\bibinfo {author} {\bibfnamefont {C.}~\bibnamefont
  {{Cichowlas}}}\ and\ \bibinfo {author} {\bibfnamefont {M.-E.}\ \bibnamefont
  {{Brachet}}},\ }\href {https://doi.org/10.1016/j.fluiddyn.2004.09.005}
  {\bibfield  {journal} {\bibinfo  {journal} {Fluid Dyn. Res.}\ }\textbf
  {\bibinfo {volume} {36}},\ \bibinfo {pages} {239} (\bibinfo {year}
  {2005})}\BibitemShut {NoStop}%
\bibitem [{\citenamefont {{Kolluru}}\ \emph {et~al.}(2022)\citenamefont
  {{Kolluru}}, \citenamefont {{Sharma}},\ and\ \citenamefont
  {{Pandit}}}]{Kolluru2022PhRvE}%
  \BibitemOpen
  \bibfield  {author} {\bibinfo {author} {\bibfnamefont {S.~S.~V.}\
  \bibnamefont {{Kolluru}}}, \bibinfo {author} {\bibfnamefont {P.}~\bibnamefont
  {{Sharma}}},\ and\ \bibinfo {author} {\bibfnamefont {R.}~\bibnamefont
  {{Pandit}}},\ }\href {https://doi.org/10.1103/PhysRevE.105.065107} {\bibfield
   {journal} {\bibinfo  {journal} {\pre}\ }\textbf {\bibinfo {volume} {105}},\
  \bibinfo {eid} {065107} (\bibinfo {year} {2022})},\ \Eprint
  {https://arxiv.org/abs/2012.14182} {arXiv:2012.14182} \BibitemShut {NoStop}%
\bibitem [{\citenamefont {Ray}\ \emph {et~al.}(2011)\citenamefont {Ray},
  \citenamefont {Frisch}, \citenamefont {Nazarenko},\ and\ \citenamefont
  {Matsumoto}}]{Ray2011}%
  \BibitemOpen
  \bibfield  {author} {\bibinfo {author} {\bibfnamefont {S.~S.}\ \bibnamefont
  {Ray}}, \bibinfo {author} {\bibfnamefont {U.}~\bibnamefont {Frisch}},
  \bibinfo {author} {\bibfnamefont {S.}~\bibnamefont {Nazarenko}},\ and\
  \bibinfo {author} {\bibfnamefont {T.}~\bibnamefont {Matsumoto}},\ }\href
  {https://doi.org/10.1103/PhysRevE.84.016301} {\bibfield  {journal} {\bibinfo
  {journal} {Phys. Rev. E}\ }\textbf {\bibinfo {volume} {84}},\ \bibinfo
  {pages} {016301} (\bibinfo {year} {2011})},\ \Eprint
  {https://arxiv.org/abs/1011.1826} {arXiv:1011.1826} \BibitemShut {NoStop}%
\bibitem [{\citenamefont {{Venkataraman}}\ and\ \citenamefont {{Sankar
  Ray}}(2017)}]{VenkataramanRay2017RSPSA.47360585V}%
  \BibitemOpen
  \bibfield  {author} {\bibinfo {author} {\bibfnamefont {D.}~\bibnamefont
  {{Venkataraman}}}\ and\ \bibinfo {author} {\bibfnamefont {S.}~\bibnamefont
  {{Sankar Ray}}},\ }\href {https://doi.org/10.1098/rspa.2016.0585} {\bibfield
  {journal} {\bibinfo  {journal} {Proc. R. Soc. Lond. A}\ }\textbf {\bibinfo
  {volume} {473}},\ \bibinfo {eid} {20160585} (\bibinfo {year} {2017})},\
  \Eprint {https://arxiv.org/abs/1608.07574} {arXiv:1608.07574} \BibitemShut
  {NoStop}%
\bibitem [{\citenamefont {{Clark Di Leoni}}\ \emph {et~al.}(2018)\citenamefont
  {{Clark Di Leoni}}, \citenamefont {{Mininni}},\ and\ \citenamefont
  {{Brachet}}}]{Leoni2018PhRvF...3a4603C}%
  \BibitemOpen
  \bibfield  {author} {\bibinfo {author} {\bibfnamefont {P.}~\bibnamefont
  {{Clark Di Leoni}}}, \bibinfo {author} {\bibfnamefont {P.~D.}\ \bibnamefont
  {{Mininni}}},\ and\ \bibinfo {author} {\bibfnamefont {M.~E.}\ \bibnamefont
  {{Brachet}}},\ }\href {https://doi.org/10.1103/PhysRevFluids.3.014603}
  {\bibfield  {journal} {\bibinfo  {journal} {Phys. Rev. Fluids}\ }\textbf
  {\bibinfo {volume} {3}},\ \bibinfo {eid} {014603} (\bibinfo {year} {2018})},\
  \Eprint {https://arxiv.org/abs/1711.08618} {arXiv:1711.08618} \BibitemShut
  {NoStop}%
\bibitem [{\citenamefont {{Banerjee}}\ and\ \citenamefont
  {{Ray}}(2014)}]{2014Banerjee}%
  \BibitemOpen
  \bibfield  {author} {\bibinfo {author} {\bibfnamefont {D.}~\bibnamefont
  {{Banerjee}}}\ and\ \bibinfo {author} {\bibfnamefont {S.~S.}\ \bibnamefont
  {{Ray}}},\ }\href {https://doi.org/10.1103/PhysRevE.90.041001} {\bibfield
  {journal} {\bibinfo  {journal} {\pre}\ }\textbf {\bibinfo {volume} {90}},\
  \bibinfo {eid} {041001} (\bibinfo {year} {2014})},\ \Eprint
  {https://arxiv.org/abs/1403.6599} {arXiv:1403.6599} \BibitemShut {NoStop}%
\bibitem [{\citenamefont {{Murugan}}\ and\ \citenamefont
  {{Ray}}(2023)}]{2023Murugan}%
  \BibitemOpen
  \bibfield  {author} {\bibinfo {author} {\bibfnamefont {S.~D.}\ \bibnamefont
  {{Murugan}}}\ and\ \bibinfo {author} {\bibfnamefont {S.~S.}\ \bibnamefont
  {{Ray}}},\ }\href {https://doi.org/10.1103/PhysRevFluids.8.084605} {\bibfield
   {journal} {\bibinfo  {journal} {Phys. Rev. Fluids}\ }\textbf {\bibinfo
  {volume} {8}},\ \bibinfo {eid} {084605} (\bibinfo {year} {2023})},\ \Eprint
  {https://arxiv.org/abs/2209.05046} {arXiv:2209.05046} \BibitemShut {NoStop}%
\bibitem [{\citenamefont {{S. Venkata Kolluru}}\ \emph
  {et~al.}(2024)\citenamefont {{S. Venkata Kolluru}}, \citenamefont {{Besse}},\
  and\ \citenamefont {{Pandit}}}]{KolluruBessePandit2024}%
  \BibitemOpen
  \bibfield  {author} {\bibinfo {author} {\bibfnamefont {S.}~\bibnamefont {{S.
  Venkata Kolluru}}}, \bibinfo {author} {\bibfnamefont {N.}~\bibnamefont
  {{Besse}}},\ and\ \bibinfo {author} {\bibfnamefont {R.}~\bibnamefont
  {{Pandit}}},\ }\href
  {https://doi.org/https://doi.org/10.1016/j.jcp.2024.113446} {\bibfield
  {journal} {\bibinfo  {journal} {J. Comput. Phys.}\ }\textbf {\bibinfo
  {volume} {519}},\ \bibinfo {pages} {113446} (\bibinfo {year} {2024})},\
  \Eprint {https://arxiv.org/abs/2402.17688} {arXiv:2402.17688} \BibitemShut
  {NoStop}%
\bibitem [{\citenamefont {{Taylor}}\ and\ \citenamefont
  {{Green}}(1937)}]{Taylor-Green-1937RSPSA.158..499T}%
  \BibitemOpen
  \bibfield  {author} {\bibinfo {author} {\bibfnamefont {G.~I.}\ \bibnamefont
  {{Taylor}}}\ and\ \bibinfo {author} {\bibfnamefont {A.~E.}\ \bibnamefont
  {{Green}}},\ }\href {https://doi.org/10.1098/rspa.1937.0036} {\bibfield
  {journal} {\bibinfo  {journal} {Proc. R. Soc. A}\ }\textbf {\bibinfo {volume}
  {158}},\ \bibinfo {pages} {499} (\bibinfo {year} {1937})}\BibitemShut
  {NoStop}%
\bibitem [{\citenamefont {{van Dyke}}(1974)}]{vanDyke1974}%
  \BibitemOpen
  \bibfield  {author} {\bibinfo {author} {\bibfnamefont {M.}~\bibnamefont {{van
  Dyke}}},\ }\href {https://doi.org/10.1093/qjmam/27.4.423} {\bibfield
  {journal} {\bibinfo  {journal} {Q. Jl Mech. appl. Math.}\ }\textbf {\bibinfo
  {volume} {27}},\ \bibinfo {pages} {423} (\bibinfo {year} {1974})}\BibitemShut
  {NoStop}%
\bibitem [{\citenamefont {{van Dyke}}(1975)}]{Dyke-1975SJAM...28..720V}%
  \BibitemOpen
  \bibfield  {author} {\bibinfo {author} {\bibfnamefont {M.}~\bibnamefont {{van
  Dyke}}},\ }\href {https://doi.org/10.1137/0128060} {\bibfield  {journal}
  {\bibinfo  {journal} {SIAM J. Appl. Math.}\ }\textbf {\bibinfo {volume}
  {28}},\ \bibinfo {pages} {720} (\bibinfo {year} {1975})}\BibitemShut
  {NoStop}%
\bibitem [{\citenamefont {{Morf}}\ \emph {et~al.}(1980)\citenamefont {{Morf}},
  \citenamefont {{Orszag}},\ and\ \citenamefont {{Frisch}}}]{Morf1980}%
  \BibitemOpen
  \bibfield  {author} {\bibinfo {author} {\bibfnamefont {R.~H.}\ \bibnamefont
  {{Morf}}}, \bibinfo {author} {\bibfnamefont {S.~A.}\ \bibnamefont
  {{Orszag}}},\ and\ \bibinfo {author} {\bibfnamefont {U.}~\bibnamefont
  {{Frisch}}},\ }\href {https://doi.org/10.1103/PhysRevLett.44.572} {\bibfield
  {journal} {\bibinfo  {journal} {\prl}\ }\textbf {\bibinfo {volume} {44}},\
  \bibinfo {pages} {572} (\bibinfo {year} {1980})}\BibitemShut {NoStop}%
\bibitem [{\citenamefont {{Brachet}}\ \emph
  {et~al.}(1983{\natexlab{b}})\citenamefont {{Brachet}}, \citenamefont
  {{Meiron}}, \citenamefont {{Orszag}}, \citenamefont {{Nickel}}, \citenamefont
  {{Morf}},\ and\ \citenamefont {{Frisch}}}]{1983Brachet}%
  \BibitemOpen
  \bibfield  {author} {\bibinfo {author} {\bibfnamefont {M.~E.}\ \bibnamefont
  {{Brachet}}}, \bibinfo {author} {\bibfnamefont {D.~I.}\ \bibnamefont
  {{Meiron}}}, \bibinfo {author} {\bibfnamefont {S.~A.}\ \bibnamefont
  {{Orszag}}}, \bibinfo {author} {\bibfnamefont {B.~G.}\ \bibnamefont
  {{Nickel}}}, \bibinfo {author} {\bibfnamefont {R.~H.}\ \bibnamefont
  {{Morf}}},\ and\ \bibinfo {author} {\bibfnamefont {U.}~\bibnamefont
  {{Frisch}}},\ }\href {https://doi.org/10.1017/S0022112083001159} {\bibfield
  {journal} {\bibinfo  {journal} {J. Fluid Mech.}\ }\textbf {\bibinfo {volume}
  {130}},\ \bibinfo {pages} {411} (\bibinfo {year}
  {1983}{\natexlab{b}})}\BibitemShut {NoStop}%
\bibitem [{\citenamefont {{Brachet}}\ \emph {et~al.}(1984)\citenamefont
  {{Brachet}}, \citenamefont {{Meiron}}, \citenamefont {{Orszag}},
  \citenamefont {{Nickel}}, \citenamefont {{Morf}},\ and\ \citenamefont
  {{Frisch}}}]{1984Brachet}%
  \BibitemOpen
  \bibfield  {author} {\bibinfo {author} {\bibfnamefont {M.~E.}\ \bibnamefont
  {{Brachet}}}, \bibinfo {author} {\bibfnamefont {D.}~\bibnamefont {{Meiron}}},
  \bibinfo {author} {\bibfnamefont {S.}~\bibnamefont {{Orszag}}}, \bibinfo
  {author} {\bibfnamefont {B.}~\bibnamefont {{Nickel}}}, \bibinfo {author}
  {\bibfnamefont {R.}~\bibnamefont {{Morf}}},\ and\ \bibinfo {author}
  {\bibfnamefont {U.}~\bibnamefont {{Frisch}}},\ }\href
  {https://doi.org/10.1007/BF01009458} {\bibfield  {journal} {\bibinfo
  {journal} {J. Stat. Phys.}\ }\textbf {\bibinfo {volume} {34}},\ \bibinfo
  {pages} {1049} (\bibinfo {year} {1984})}\BibitemShut {NoStop}%
\bibitem [{\citenamefont {{Pelz}}\ and\ \citenamefont
  {{Gulak}}(1997)}]{1997Pelz}%
  \BibitemOpen
  \bibfield  {author} {\bibinfo {author} {\bibfnamefont {R.~B.}\ \bibnamefont
  {{Pelz}}}\ and\ \bibinfo {author} {\bibfnamefont {Y.}~\bibnamefont
  {{Gulak}}},\ }\href {https://doi.org/10.1103/PhysRevLett.79.4998} {\bibfield
  {journal} {\bibinfo  {journal} {\prl}\ }\textbf {\bibinfo {volume} {79}},\
  \bibinfo {pages} {4998} (\bibinfo {year} {1997})}\BibitemShut {NoStop}%
\bibitem [{\citenamefont {{Morf}}\ \emph {et~al.}(1981)\citenamefont {{Morf}},
  \citenamefont {{Orszag}}, \citenamefont {{Meiron}}, \citenamefont
  {{Meneguzzi}},\ and\ \citenamefont {{Frisch}}}]{1981Morf}%
  \BibitemOpen
  \bibfield  {author} {\bibinfo {author} {\bibfnamefont {R.~H.}\ \bibnamefont
  {{Morf}}}, \bibinfo {author} {\bibfnamefont {S.~A.}\ \bibnamefont
  {{Orszag}}}, \bibinfo {author} {\bibfnamefont {D.~I.}\ \bibnamefont
  {{Meiron}}}, \bibinfo {author} {\bibfnamefont {M.}~\bibnamefont
  {{Meneguzzi}}},\ and\ \bibinfo {author} {\bibfnamefont {U.}~\bibnamefont
  {{Frisch}}},\ }in\ \href {https://doi.org/10.1007/3-540-10694-4_44} {\emph
  {\bibinfo {booktitle} {Numerical Methods in Fluid Dynamics}}},\ Vol.\
  \bibinfo {volume} {141},\ \bibinfo {editor} {edited by\ \bibinfo {editor}
  {\bibfnamefont {W.~C.}\ \bibnamefont {{Reynolds}}}\ and\ \bibinfo {editor}
  {\bibfnamefont {R.~W.}\ \bibnamefont {{MacCormack}}}}\ (\bibinfo  {publisher}
  {Springer},\ \bibinfo {year} {1981})\ pp.\ \bibinfo {pages}
  {292--298}\BibitemShut {NoStop}%
\bibitem [{\citenamefont {{Frisch}}\ and\ \citenamefont
  {{Morf}}(1981)}]{frisch-morf}%
  \BibitemOpen
  \bibfield  {author} {\bibinfo {author} {\bibfnamefont {U.}~\bibnamefont
  {{Frisch}}}\ and\ \bibinfo {author} {\bibfnamefont {R.}~\bibnamefont
  {{Morf}}},\ }\href {https://doi.org/10.1103/PhysRevA.23.2673} {\bibfield
  {journal} {\bibinfo  {journal} {Phys. Rev. A}\ }\textbf {\bibinfo {volume}
  {23}},\ \bibinfo {pages} {2673} (\bibinfo {year} {1981})}\BibitemShut
  {NoStop}%
\bibitem [{\citenamefont {{Meiron}}\ \emph {et~al.}(1982)\citenamefont
  {{Meiron}}, \citenamefont {{Baker}},\ and\ \citenamefont
  {{Orszag}}}]{1982Meiron}%
  \BibitemOpen
  \bibfield  {author} {\bibinfo {author} {\bibfnamefont {D.~I.}\ \bibnamefont
  {{Meiron}}}, \bibinfo {author} {\bibfnamefont {G.~R.}\ \bibnamefont
  {{Baker}}},\ and\ \bibinfo {author} {\bibfnamefont {S.~A.}\ \bibnamefont
  {{Orszag}}},\ }\href {https://doi.org/10.1017/S0022112082000159} {\bibfield
  {journal} {\bibinfo  {journal} {J. Fluid Mech.}\ }\textbf {\bibinfo {volume}
  {114}},\ \bibinfo {pages} {283} (\bibinfo {year} {1982})}\BibitemShut
  {NoStop}%
\bibitem [{\citenamefont {{Ohkitani}}\ and\ \citenamefont
  {{Gibbon}}(2000)}]{2000Ohkitani}%
  \BibitemOpen
  \bibfield  {author} {\bibinfo {author} {\bibfnamefont {K.}~\bibnamefont
  {{Ohkitani}}}\ and\ \bibinfo {author} {\bibfnamefont {J.~D.}\ \bibnamefont
  {{Gibbon}}},\ }\href {https://doi.org/10.1063/1.1321256} {\bibfield
  {journal} {\bibinfo  {journal} {Phys. Fluids}\ }\textbf {\bibinfo {volume}
  {12}},\ \bibinfo {pages} {3181} (\bibinfo {year} {2000})}\BibitemShut
  {NoStop}%
\bibitem [{\citenamefont {Rampf}\ \emph {et~al.}(2022)\citenamefont {Rampf},
  \citenamefont {Frisch},\ and\ \citenamefont {Hahn}}]{Rampf:2022apj}%
  \BibitemOpen
  \bibfield  {author} {\bibinfo {author} {\bibfnamefont {C.}~\bibnamefont
  {Rampf}}, \bibinfo {author} {\bibfnamefont {U.}~\bibnamefont {Frisch}},\ and\
  \bibinfo {author} {\bibfnamefont {O.}~\bibnamefont {Hahn}},\ }\href
  {https://doi.org/10.1103/PhysRevFluids.7.104610} {\bibfield  {journal}
  {\bibinfo  {journal} {Phys. Rev. Fluids}\ }\textbf {\bibinfo {volume} {7}},\
  \bibinfo {pages} {104610} (\bibinfo {year} {2022})},\ \Eprint
  {https://arxiv.org/abs/2207.12416} {arXiv:2207.12416} \BibitemShut {NoStop}%
\bibitem [{\citenamefont {Choi}\ \emph {et~al.}(2017)\citenamefont {Choi},
  \citenamefont {Hou}, \citenamefont {Kiselev}, \citenamefont {Luo},
  \citenamefont {Sverak},\ and\ \citenamefont {Yao}}]{Choi2017}%
  \BibitemOpen
  \bibfield  {author} {\bibinfo {author} {\bibfnamefont {K.}~\bibnamefont
  {Choi}}, \bibinfo {author} {\bibfnamefont {T.~Y.}\ \bibnamefont {Hou}},
  \bibinfo {author} {\bibfnamefont {A.}~\bibnamefont {Kiselev}}, \bibinfo
  {author} {\bibfnamefont {G.}~\bibnamefont {Luo}}, \bibinfo {author}
  {\bibfnamefont {V.}~\bibnamefont {Sverak}},\ and\ \bibinfo {author}
  {\bibfnamefont {Y.}~\bibnamefont {Yao}},\ }\href
  {https://doi.org/https://doi.org/10.1002/cpa.21697} {\bibfield  {journal}
  {\bibinfo  {journal} {Commun. Pure Appl. Math.}\ }\textbf {\bibinfo {volume}
  {70}},\ \bibinfo {pages} {2218} (\bibinfo {year} {2017})},\ \Eprint
  {https://arxiv.org/abs/1407.4776} {arXiv:1407.4776} \BibitemShut {NoStop}%
\bibitem [{\citenamefont {{Huang}}\ \emph {et~al.}(2025)\citenamefont
  {{Huang}}, \citenamefont {{Qin}}, \citenamefont {{Wang}},\ and\ \citenamefont
  {{Wei}}}]{Huang2025}%
  \BibitemOpen
  \bibfield  {author} {\bibinfo {author} {\bibfnamefont {D.}~\bibnamefont
  {{Huang}}}, \bibinfo {author} {\bibfnamefont {X.}~\bibnamefont {{Qin}}},
  \bibinfo {author} {\bibfnamefont {X.}~\bibnamefont {{Wang}}},\ and\ \bibinfo
  {author} {\bibfnamefont {D.}~\bibnamefont {{Wei}}},\ }\href
  {https://doi.org/10.1007/s00220-025-05429-9} {\bibfield  {journal} {\bibinfo
  {journal} {Commun. Math. Phys.}\ }\textbf {\bibinfo {volume} {406}},\
  \bibinfo {eid} {243} (\bibinfo {year} {2025})},\ \Eprint
  {https://arxiv.org/abs/2308.01528} {arXiv:2308.01528} \BibitemShut {NoStop}%
\bibitem [{\citenamefont {{Do}}\ \emph {et~al.}(2018)\citenamefont {{Do}},
  \citenamefont {{Kiselev}},\ and\ \citenamefont {{Xu}}}]{2018Do}%
  \BibitemOpen
  \bibfield  {author} {\bibinfo {author} {\bibfnamefont {T.}~\bibnamefont
  {{Do}}}, \bibinfo {author} {\bibfnamefont {A.}~\bibnamefont {{Kiselev}}},\
  and\ \bibinfo {author} {\bibfnamefont {X.}~\bibnamefont {{Xu}}},\ }\href
  {https://doi.org/10.1007/s00332-016-9340-7} {\bibfield  {journal} {\bibinfo
  {journal} {J. Nonlinear Sci.}\ }\textbf {\bibinfo {volume} {28}},\ \bibinfo
  {pages} {2127} (\bibinfo {year} {2018})},\ \Eprint
  {https://arxiv.org/abs/1604.07118} {arXiv:1604.07118} \BibitemShut {NoStop}%
\bibitem [{\citenamefont {Venkata~Kolluru}\ and\ \citenamefont
  {Pandit}(2024)}]{KolluruPandit2024}%
  \BibitemOpen
  \bibfield  {author} {\bibinfo {author} {\bibfnamefont {S.~S.}\ \bibnamefont
  {Venkata~Kolluru}}\ and\ \bibinfo {author} {\bibfnamefont {R.}~\bibnamefont
  {Pandit}},\ }\href {https://doi.org/10.1063/5.0222257} {\bibfield  {journal}
  {\bibinfo  {journal} {Phys. Fluids}\ }\textbf {\bibinfo {volume} {36}},\
  \bibinfo {pages} {097159} (\bibinfo {year} {2024})},\ \Eprint
  {https://arxiv.org/abs/2406.04228} {arXiv:2406.04228} \BibitemShut {NoStop}%
\bibitem [{\citenamefont {{Constantin}}\ \emph {et~al.}(1985)\citenamefont
  {{Constantin}}, \citenamefont {{Lax}},\ and\ \citenamefont
  {{Majda}}}]{ConstantinLaxMajda1985}%
  \BibitemOpen
  \bibfield  {author} {\bibinfo {author} {\bibfnamefont {P.}~\bibnamefont
  {{Constantin}}}, \bibinfo {author} {\bibfnamefont {P.~D.}\ \bibnamefont
  {{Lax}}},\ and\ \bibinfo {author} {\bibfnamefont {A.}~\bibnamefont
  {{Majda}}},\ }\href {https://doi.org/10.1002/cpa.3160380605} {\bibfield
  {journal} {\bibinfo  {journal} {Commun. Pure Appl. Math.}\ }\textbf {\bibinfo
  {volume} {38}},\ \bibinfo {pages} {715} (\bibinfo {year} {1985})}\BibitemShut
  {NoStop}%
\bibitem [{\citenamefont {Zheligovsky}\ and\ \citenamefont
  {Frisch}(2014)}]{Zheligovsky:2013eca}%
  \BibitemOpen
  \bibfield  {author} {\bibinfo {author} {\bibfnamefont {V.}~\bibnamefont
  {Zheligovsky}}\ and\ \bibinfo {author} {\bibfnamefont {U.}~\bibnamefont
  {Frisch}},\ }\href {https://doi.org/10.1017/jfm.2014.221} {\bibfield
  {journal} {\bibinfo  {journal} {J. Fluid Mech.}\ }\textbf {\bibinfo {volume}
  {749}},\ \bibinfo {pages} {404} (\bibinfo {year} {2014})},\ \Eprint
  {https://arxiv.org/abs/1312.6320} {arXiv:1312.6320} \BibitemShut {NoStop}%
\bibitem [{\citenamefont {{Shukla}}\ \emph {et~al.}(2013)\citenamefont
  {{Shukla}}, \citenamefont {{Brachet}},\ and\ \citenamefont
  {{Pandit}}}]{2013NJPh...15k3025S}%
  \BibitemOpen
  \bibfield  {author} {\bibinfo {author} {\bibfnamefont {V.}~\bibnamefont
  {{Shukla}}}, \bibinfo {author} {\bibfnamefont {M.}~\bibnamefont
  {{Brachet}}},\ and\ \bibinfo {author} {\bibfnamefont {R.}~\bibnamefont
  {{Pandit}}},\ }\href {https://doi.org/10.1088/1367-2630/15/11/113025}
  {\bibfield  {journal} {\bibinfo  {journal} {New J. Phys.}\ }\textbf {\bibinfo
  {volume} {15}},\ \bibinfo {eid} {113025} (\bibinfo {year} {2013})},\ \Eprint
  {https://arxiv.org/abs/1301.3383} {arXiv:1301.3383} \BibitemShut {NoStop}%
\bibitem [{\citenamefont {{Podvigina}}\ \emph {et~al.}(2016)\citenamefont
  {{Podvigina}}, \citenamefont {{Zheligovsky}},\ and\ \citenamefont
  {{Frisch}}}]{Podvigina2016}%
  \BibitemOpen
  \bibfield  {author} {\bibinfo {author} {\bibfnamefont {O.}~\bibnamefont
  {{Podvigina}}}, \bibinfo {author} {\bibfnamefont {V.}~\bibnamefont
  {{Zheligovsky}}},\ and\ \bibinfo {author} {\bibfnamefont {U.}~\bibnamefont
  {{Frisch}}},\ }\href {https://doi.org/10.1016/j.jcp.2015.11.045} {\bibfield
  {journal} {\bibinfo  {journal} {J. Comput. Phys.}\ }\textbf {\bibinfo
  {volume} {306}},\ \bibinfo {pages} {320} (\bibinfo {year} {2016})},\ \Eprint
  {https://arxiv.org/abs/1504.05030} {arXiv:1504.05030} \BibitemShut {NoStop}%
\bibitem [{\citenamefont {{Murugan}}\ \emph {et~al.}(2020)\citenamefont
  {{Murugan}}, \citenamefont {{Frisch}}, \citenamefont {{Nazarenko}},
  \citenamefont {{Besse}},\ and\ \citenamefont {{Ray}}}]{2020PhRvR...2c3202M}%
  \BibitemOpen
  \bibfield  {author} {\bibinfo {author} {\bibfnamefont {S.~D.}\ \bibnamefont
  {{Murugan}}}, \bibinfo {author} {\bibfnamefont {U.}~\bibnamefont {{Frisch}}},
  \bibinfo {author} {\bibfnamefont {S.}~\bibnamefont {{Nazarenko}}}, \bibinfo
  {author} {\bibfnamefont {N.}~\bibnamefont {{Besse}}},\ and\ \bibinfo {author}
  {\bibfnamefont {S.~S.}\ \bibnamefont {{Ray}}},\ }\href
  {https://doi.org/10.1103/PhysRevResearch.2.033202} {\bibfield  {journal}
  {\bibinfo  {journal} {Phys. Rev. Res.}\ }\textbf {\bibinfo {volume} {2}},\
  \bibinfo {eid} {033202} (\bibinfo {year} {2020})},\ \Eprint
  {https://arxiv.org/abs/2001.04819} {arXiv:2001.04819} \BibitemShut {NoStop}%
\bibitem [{\citenamefont {Mercer}\ and\ \citenamefont
  {Roberts}(1990)}]{MercerRoberts}%
  \BibitemOpen
  \bibfield  {author} {\bibinfo {author} {\bibfnamefont {G.~N.}\ \bibnamefont
  {Mercer}}\ and\ \bibinfo {author} {\bibfnamefont {A.~J.}\ \bibnamefont
  {Roberts}},\ }\href {https://doi.org/10.1137/0150091} {\bibfield  {journal}
  {\bibinfo  {journal} {SIAM J. Appl. Math.}\ }\textbf {\bibinfo {volume}
  {50}},\ \bibinfo {pages} {1547} (\bibinfo {year} {1990})}\BibitemShut
  {NoStop}%
\bibitem [{\citenamefont {{Domb}}\ and\ \citenamefont
  {{Sykes}}(1957)}]{DombSykes1957}%
  \BibitemOpen
  \bibfield  {author} {\bibinfo {author} {\bibfnamefont {C.}~\bibnamefont
  {{Domb}}}\ and\ \bibinfo {author} {\bibfnamefont {M.~F.}\ \bibnamefont
  {{Sykes}}},\ }\href {https://doi.org/10.1098/rspa.1957.0078} {\bibfield
  {journal} {\bibinfo  {journal} {Proc. R. Soc. Lond. A}\ }\textbf {\bibinfo
  {volume} {240}},\ \bibinfo {pages} {214} (\bibinfo {year}
  {1957})}\BibitemShut {NoStop}%
\bibitem [{\citenamefont {Rampf}\ and\ \citenamefont
  {Hahn}(2023)}]{Rampf:2022eiu}%
  \BibitemOpen
  \bibfield  {author} {\bibinfo {author} {\bibfnamefont {C.}~\bibnamefont
  {Rampf}}\ and\ \bibinfo {author} {\bibfnamefont {O.}~\bibnamefont {Hahn}},\
  }\href {https://doi.org/10.1103/PhysRevD.107.023515} {\bibfield  {journal}
  {\bibinfo  {journal} {Phys. Rev. D}\ }\textbf {\bibinfo {volume} {107}},\
  \bibinfo {pages} {023515} (\bibinfo {year} {2023})},\ \Eprint
  {https://arxiv.org/abs/2211.02053} {arXiv:2211.02053} \BibitemShut {NoStop}%
\bibitem [{\citenamefont {Rampf}\ \emph {et~al.}(2023)\citenamefont {Rampf},
  \citenamefont {Saga}, \citenamefont {Taruya},\ and\ \citenamefont
  {Colombi}}]{Rampf:2023fzq}%
  \BibitemOpen
  \bibfield  {author} {\bibinfo {author} {\bibfnamefont {C.}~\bibnamefont
  {Rampf}}, \bibinfo {author} {\bibfnamefont {S.}~\bibnamefont {Saga}},
  \bibinfo {author} {\bibfnamefont {A.}~\bibnamefont {Taruya}},\ and\ \bibinfo
  {author} {\bibfnamefont {S.}~\bibnamefont {Colombi}},\ }\href
  {https://doi.org/10.1103/PhysRevD.108.103513} {\bibfield  {journal} {\bibinfo
   {journal} {Phys. Rev. D}\ }\textbf {\bibinfo {volume} {108}},\ \bibinfo
  {pages} {103513} (\bibinfo {year} {2023})},\ \Eprint
  {https://arxiv.org/abs/2303.12832} {arXiv:2303.12832} \BibitemShut {NoStop}%
\bibitem [{\citenamefont {Rampf}\ and\ \citenamefont
  {Frisch}(2017)}]{Rampf:2017jan}%
  \BibitemOpen
  \bibfield  {author} {\bibinfo {author} {\bibfnamefont {C.}~\bibnamefont
  {Rampf}}\ and\ \bibinfo {author} {\bibfnamefont {U.}~\bibnamefont {Frisch}},\
  }\href {https://doi.org/10.1093/mnras/stx1613} {\bibfield  {journal}
  {\bibinfo  {journal} {Mon. Not. Roy. Astron. Soc.}\ }\textbf {\bibinfo
  {volume} {471}},\ \bibinfo {pages} {671} (\bibinfo {year} {2017})},\ \Eprint
  {https://arxiv.org/abs/1705.08456} {arXiv:1705.08456} \BibitemShut {NoStop}%
\bibitem [{\citenamefont {Rampf}(2021)}]{Rampf:2021rqu}%
  \BibitemOpen
  \bibfield  {author} {\bibinfo {author} {\bibfnamefont {C.}~\bibnamefont
  {Rampf}},\ }\href {https://doi.org/10.1007/s41614-021-00055-z} {\bibfield
  {journal} {\bibinfo  {journal} {Rev. Mod. Plasma Phys.}\ }\textbf {\bibinfo
  {volume} {5}},\ \bibinfo {pages} {10} (\bibinfo {year} {2021})},\ \Eprint
  {https://arxiv.org/abs/2110.06265} {arXiv:2110.06265} \BibitemShut {NoStop}%
\bibitem [{\citenamefont {{Ferrari}}(1993)}]{1993CMaPh.155..277F}%
  \BibitemOpen
  \bibfield  {author} {\bibinfo {author} {\bibfnamefont {A.~B.}\ \bibnamefont
  {{Ferrari}}},\ }\href {https://doi.org/10.1007/BF02097394} {\bibfield
  {journal} {\bibinfo  {journal} {Commun. Math. Phys.}\ }\textbf {\bibinfo
  {volume} {155}},\ \bibinfo {pages} {277} (\bibinfo {year}
  {1993})}\BibitemShut {NoStop}%
\bibitem [{\citenamefont {{Matsumoto}}\ \emph {et~al.}(2008)\citenamefont
  {{Matsumoto}}, \citenamefont {{Bec}},\ and\ \citenamefont
  {{Frisch}}}]{2008PhyD..237.1951M}%
  \BibitemOpen
  \bibfield  {author} {\bibinfo {author} {\bibfnamefont {T.}~\bibnamefont
  {{Matsumoto}}}, \bibinfo {author} {\bibfnamefont {J.}~\bibnamefont {{Bec}}},\
  and\ \bibinfo {author} {\bibfnamefont {U.}~\bibnamefont {{Frisch}}},\ }\href
  {https://doi.org/10.1016/j.physd.2007.11.007} {\bibfield  {journal} {\bibinfo
   {journal} {Physica D.}\ }\textbf {\bibinfo {volume} {237}},\ \bibinfo
  {pages} {1951} (\bibinfo {year} {2008})},\ \Eprint
  {https://arxiv.org/abs/0709.0219} {arXiv:0709.0219} \BibitemShut {NoStop}%
\bibitem [{\citenamefont {{Besse}}\ and\ \citenamefont
  {{Frisch}}(2017)}]{Besse2017}%
  \BibitemOpen
  \bibfield  {author} {\bibinfo {author} {\bibfnamefont {N.}~\bibnamefont
  {{Besse}}}\ and\ \bibinfo {author} {\bibfnamefont {U.}~\bibnamefont
  {{Frisch}}},\ }\href {https://doi.org/10.1007/s00220-016-2816-3} {\bibfield
  {journal} {\bibinfo  {journal} {Comm. Math. Phys.}\ }\textbf {\bibinfo
  {volume} {351}},\ \bibinfo {pages} {689} (\bibinfo {year} {2017})},\ \Eprint
  {https://arxiv.org/abs/1603.09219} {arXiv:1603.09219} \BibitemShut {NoStop}%
\bibitem [{\citenamefont {{Hertel}}\ \emph {et~al.}(2022)\citenamefont
  {{Hertel}}, \citenamefont {{Besse}},\ and\ \citenamefont
  {{Frisch}}}]{Hertel2022}%
  \BibitemOpen
  \bibfield  {author} {\bibinfo {author} {\bibfnamefont {T.}~\bibnamefont
  {{Hertel}}}, \bibinfo {author} {\bibfnamefont {N.}~\bibnamefont {{Besse}}},\
  and\ \bibinfo {author} {\bibfnamefont {U.}~\bibnamefont {{Frisch}}},\ }\href
  {https://doi.org/10.1016/j.jcp.2021.110758} {\bibfield  {journal} {\bibinfo
  {journal} {J. Comput. Phys.}\ }\textbf {\bibinfo {volume} {449}},\ \bibinfo
  {eid} {110758} (\bibinfo {year} {2022})}\BibitemShut {NoStop}%
\bibitem [{\citenamefont {Elgindi}\ and\ \citenamefont
  {Jeong}(2020)}]{ElgindiJeong2020}%
  \BibitemOpen
  \bibfield  {author} {\bibinfo {author} {\bibfnamefont {T.~M.}\ \bibnamefont
  {Elgindi}}\ and\ \bibinfo {author} {\bibfnamefont {I.-J.}\ \bibnamefont
  {Jeong}},\ }\href {https://doi.org/10.1007/s40818-020-00080-0} {\bibfield
  {journal} {\bibinfo  {journal} {Ann. PDE}\ }\textbf {\bibinfo {volume} {6}},\
  \bibinfo {pages} {5} (\bibinfo {year} {2020})},\ \Eprint
  {https://arxiv.org/abs/1708.02724} {arXiv:1708.02724} \BibitemShut {NoStop}%
\bibitem [{\citenamefont {{Chen}}\ and\ \citenamefont
  {{Hou}}(2021)}]{Chen-Hou-2021CMaPh.383.1559C}%
  \BibitemOpen
  \bibfield  {author} {\bibinfo {author} {\bibfnamefont {J.}~\bibnamefont
  {{Chen}}}\ and\ \bibinfo {author} {\bibfnamefont {T.~Y.}\ \bibnamefont
  {{Hou}}},\ }\href {https://doi.org/10.1007/s00220-021-04067-1} {\bibfield
  {journal} {\bibinfo  {journal} {Comm. Math. Phys.}\ }\textbf {\bibinfo
  {volume} {383}},\ \bibinfo {pages} {1559} (\bibinfo {year} {2021})},\ \Eprint
  {https://arxiv.org/abs/1910.00173} {arXiv:1910.00173} \BibitemShut {NoStop}%
\end{thebibliography}%

\appendix

\section{Details to the asymptotic analysis in Eulerian coordinates} \label{app:MR}

In Sec.\,\ref{sec:appMR-DS} we provide some details of the asymptotic analysis applied to the HL model, that lead to the results as shown in Fig.\,\ref{fig:sing} in the main text. In Sec.\,\ref{app:scale} we show that our methods can be used straightforwardly to predict the time of blow-up for scaled initial data.

\subsection{Asymptotic analysis and Extrapolation methods}\label{sec:appMR-DS}

In Sec.\,\ref{sec:asyEuler} we use the Mercer--Roberts estimator, which we repeat here for convenience
\begin{align}
 B_n^2(z) &=  \frac{W_{n+1} W_{n-1} - W_n^2}{W_{n} W_{n-2} - W_{n-1}^2} \,,  \label{eq:MRestimatorsRep}
\end{align}
where $W_n = W_n(z)$ are the Eulerian $\tau$-time coefficients obtained from the recursive relations~\eqref{eqs:eul_rec_hl1d}.
As argued in the main text, the estimator~\eqref{eq:MRestimatorsRep} in combination with 
\begin{align}
  B_n(z)  &= \frac{1}{R^{\MR}} \left(1  - [1+ \nu^{\MR}] \frac 1 n + O(n^{-2}) \right)  \label{eqs:MRextraRep}
\end{align}
can be used to retrieve the leading-order asymptotics of the vorticity in the HL model.  
\begin{figure}
 \centering 
\includegraphics[width=0.9\textwidth]{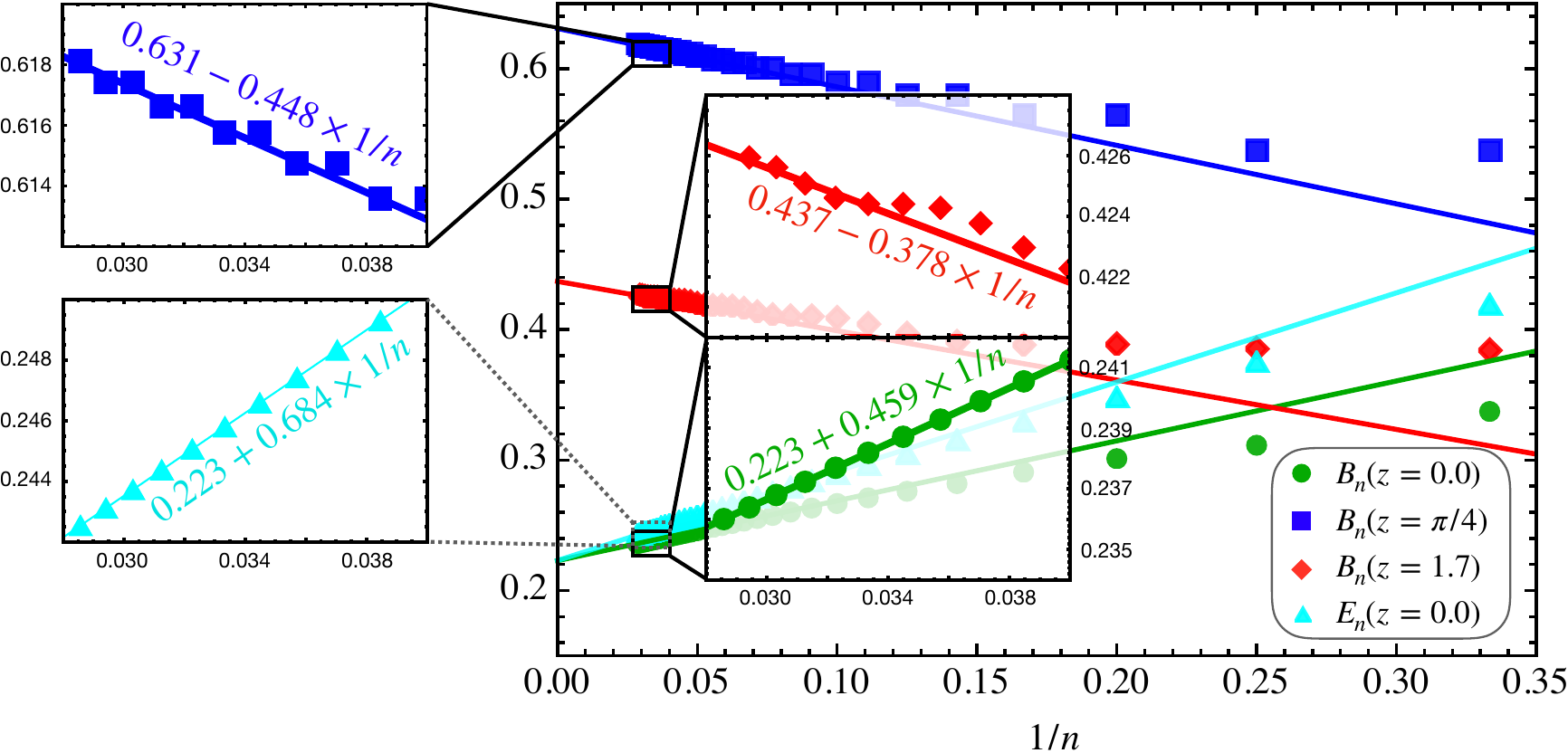}

\caption{Results of the Mercer--Roberts estimator $B_n$, evaluated at various locations. For the blow-up location~$z=0$, we also show the results of the Domb--Sykes method, with Eulerian estimator $E_n(z)= W_n(z)/W_{n-1}(z)$ (cf.\ with Lagrangian estimator given in Eq.\,\ref{eq:Dn}).
%
} \label{fig:MR-DS-Euler}
\end{figure}
For this, we show, in Fig.\,\ref{fig:MR-DS-Euler}, the $B_n$ estimator over $1/n$, evaluated at three exemplary points $z = 0.0,\pi/4, 1.7$ (green, blue and red plot markers). At the blow-up location $z=0$ (green filled circles), the~$B_n$ estimator settles into a linear behavior for sufficiently large $n = 30-35$, justifying the linear model~\eqref{eqs:MRextraRep} and thereby allowing us to retrieve the location of the singularity~$R(z=0) =: R_\star^{\MR} \simeq 4.4865$ with singularity exponent~$\nu^{\MR}(z=0) \simeq -3.060$. Regarding the location $z=\pi/4$ (blue squares), we observe the repeating pattern that $B_2 = B_3$, $B_4= B_5$, etc., together with the phase $\theta(z=\pi/4) = \pi/2$ (see left panel in Fig.\,\ref{fig:sing}). Note that an identical pattern was observed in Ref.\,\cite{Rampf:2022apj} for inviscid Burgers when the phase of the singularity is $\pi/2$ and thus perfectly aligned with the imaginary time axis. Extrapolating to the origin by using~\eqref{eqs:MRextraRep} leads to a modulus of $R(z=\pi/4) =: R_{\rm inf}^{\MR} \simeq 1.5875$, agreeing with our theoretical result to $0.013\%$ (Sec.\,\ref{sec:sing-theory}). As for the third location ($z=1.7$, red diamond marker), the trend observed for $B_n$ is not linear, even for large orders; in principle, one would need to include nonlinear terms in the asymptotic model~\eqref{eqs:MRextraRep} as done in \cite{KolluruPandit2024}.

Finally, motivated by the above findings, we also apply the Domb--Sykes (DS) test to the Eulerian vorticity at the blow-up location $z=0$, with corresponding estimator 
\be \label{eq:DnEuler}
   E_n(z=0) := \lim_{z\to 0}\frac{W_n(z)}{W_{n-1}(z)} = \frac{1}{R^{\DSE}_\star} \left(  1 - [1+\nu^{\DSE}_\star]\,\frac{1}{n}\right) 
\ee
(cf.\ Eq.\,\ref{eq:Dn} for the estimator for the Lagrangian vorticity). Crucially, this estimator assumes that the convergence-limiting singularity is located on the real time axis---as opposed to the MR estimator~$B_n$ [Eq.\,\eqref{eqs:MRextraRep}] which assumes pairs of complex-time singularities. On a practical level, $E_n$  delivers complementary predictions to $B_n$. Moreover, computing~$E_n$ for fixed $n$ requires only two Taylor coefficients, whereas $B_n$ relies on four. Thus, given that only a limited number of Taylor coefficients are available, the $E_n$ estimator is considerably more economical. 
In Fig.\,\ref{fig:MR-DS-Euler}, we show $E_n$ at the blow-up location, finding in particular $R^{\DSE}_\star \simeq 4.4842$, which agrees with the MR result to $0.049\%$, and with the Lagrangian Domb--Sykes result (Sec.\,\ref{sec:asyLag}) to~$0.018\%$.

\subsection{Scalability of extrapolation results}\label{app:scale}

Here we show that our extrapolation methods can be straightforwardly applied to predict the time of blow-up for scaled (sc) initial data of the form
\begin{align}\label{eq:ICs-scaled}
    u^{\rm sc}(z,\tau=0) =  A\sin^2 (2\pi z/L) \,, \qquad w^{\rm sc}(z,\tau=0) = 0 \,.
\end{align}
In the main text, we use $A=1$ and $L=2\pi$, respectively; Refs.~\cite{HouLuo2014,KolluruBessePandit2024} use rescaled initial data with $A=10^4$ and $L=1/6$ instead. 
Obviously, the effective rescaling  leaves the spatial location of the blow-up unaltered, but the same is not true for the blow-up time. Indeed, using~\eqref{eq:ICs-scaled} as the input of the recursive relations~\eqref{eqs:eul_rec_hl1d} and evaluating ratios of the resulting vorticity coefficients near the blow-up location ($z \to 0$), we find for the scaled coefficients
\be \label{eq:}
  \lim_{z \to 0} \frac{W_n^{\rm sc}(z)}{W_{n-1}^{\rm sc}(z)} =  \frac{2\pi A}{L} \lim_{z \to 0} \frac{W_n(z)}{W_{n-1}(z)} \stackrel{\,\text{Eq.\,}\eqref{eq:DnEuler}\,}{=} \frac{2\pi A}{L} \frac{1}{R^{\DSE}_\star} \left(  1 - [1+\nu^{\DSE}_\star]\,\frac{1}{n}\right) \,,
\ee
where $W_n^{\rm sc}(z)$ are the coefficients resulting from~\eqref{eq:ICs-scaled}, while $W_n(z)$ are those from the main text stemming from the initial data~\eqref{eq:ICs}. Thus, once the blow-up $\tau$-time is known for any set of $A$ and $L$, the corresponding blow-up time for any scaled initial data~\eqref{eq:ICs-scaled} follows readily, reading $R^{\rm sc}_\star = L  R^{\DSE}_\star/(2\pi A)$, where~$R^{\DSE}_\star \simeq 4.4842$.


\section{Code for recursive relations}\label{app:recs}

Here we provide Mathematica-12 codes for computing the time-Taylor coefficients of the 1D HL model using the Eulerian framework of Sec.\,\ref{sec:EulerianSeries} and the Lagrangian framework of Sec.\,\ref{sec:HLlagspace}.

\begin{lstlisting}[
  language=Mathematica,
  basicstyle=\ttfamily\footnotesize,
  breaklines=true,
  breakatwhitespace=true,
  caption={Code for computation of Eulerian time--Taylor coefficients in Eq.~\eqref{eqs:eul_rec_hl1d}}
]

(*----- Define parameters and functions -----*)
f[n_] = Floor[(n - 1)/2] ;   
HilbertTransform[f_, x_, X_] := Module[ {fp = FourierParameters -> {1, -1}, k}, InverseFourierTransform[ -I (2 HeavisideTheta[k]-1) *    FourierTransform[f, x, k, fp], k, X, fp]];     

(*----- Initial data -----*)
U[0, z_] := Sin[z]^2;

(*----- Recursion relations -----*)
W[n_, z_] := W[n, z] = FullSimplify[ExpToTrig[((- Sum[If[i != (n - 1 - i),  
                (V[i, z] D[W[n - 1 - i, z], z]) + (V[n - 1 - i, z] D[W[i, z], z]) , (V[i, z] D[W[i, z], z])], {i, 0, f[n], 1}] + D[U[n, z], z])/(2 n + 1) )   ]];
            
U[n_, z_] := U[n, z] = FullSimplify[ExpToTrig[((-Sum[If[i != (n - 1 - i), 
                (V[i, z] D[U[n - 1 - i, z], z]) + (V[n - 1 - i, z] D[U[i, z], z]), 
                (V[i, z] D[U[i, z], z])], {i, 0, f[n], 1}])/(2 n))]];

V[n_, z_] := V[n,z]= FullSimplify[ExpToTrig[Integrate[HilbertTransform[W[n,x],x,z],z]]];

(*----- Compute coefficients up to order nmax = 36 -----*)
coeffs = {}; nmin = 0; nmax = 36;
Do[nn = n; AppendTo[coeffs, {n, U[n, z], W[n, z], V[n, z]}], {n, nmin, nmax}];

(*----- Write to file -----*)
Fnameout =  "./ttc_Eulerian_" <> ToString[coeffs[[-1, 1]]] <> ".mx";
DumpSave[Fnameout, coeffs];
  
\end{lstlisting}
The above code is generic in the sense that it is valid for symmetric (multi-mode) initial data (see Ref.\,\cite{KolluruPandit2024} for a code handling also the asymmetric case). We remark that for special cases where the vorticity contains coefficients involving only \textit{sums} of sine or cosine modes, the computation of the Hilbert transform can be significantly accelerated by direct substitutions of the identities ${\cal H}[\sin (n z)] = - \cos (n z)$ and ${\cal H}[\cos (n z)] =  \sin (n z)$  for any positive integer~$n$. In such cases, the code should be adapted to apply these substitutions selectively, namely only when the vorticity admits this restricted functional form. Of course, similar identities for Hilbert transforms could be implemented for other (non-periodic) initial data.

\begin{lstlisting}[
  language=Mathematica,
  basicstyle=\ttfamily\footnotesize,
  caption={Code for computation of Lagrangian time-Taylor coefficients $\mPsi_n(a)$ from Eq.\,\eqref{eq:psi-rec}}
]

(*----- Define parameters and functions -----*)
fp = FourierParameters -> {1, -1}; f[n_] = Floor[(n - 1)/2] ;   

(*---- Load Eulerian output from last code -----*)
ls2 = {}; Ntmax = 36;
Fnamein = "ttc_Eulerian_" <> ToString[Ntmax] <> ".mx";
DumpGet[Fnamein]; 
Do[ U[n, z_] = coeffs[[n + 1, 2]]; W[n, z_] = coeffs[[n + 1, 3]]; 
    Ve[n, z_] = coeffs[[n + 1, 4]];  , {n, 0, Ntmax}]

(*----- Compute Eulerian fourier transform V_n(k) of V_n(z) -----*)
Do[ V[n_, z_] := V[n, z] = Simplify[ComplexExpand[TrigReduce[Ve[n, z]]]];
    Vk[n, k_] = FourierTransform[V[n, z], z, k, fp];    , {n, 0, Ntmax + 1/2} ];

(*----- Initialise to 0 -----*)
Psik[0, k_] := 0; DPsi[0, a_] := 0; Psi[0, a_] := 0;

(*------- RECURSIONS -------*)

(*--- Equation 19  ---*)
Psik[n_, k_] := Psik[n, k] = Simplify[Tk[n, k]*1/n];

(*--- Psi  ---*)
Psi[n_, a_] := Psi[n, a] = ComplexExpand[TrigReduce[ If[n < 27, FullSimplify[InverseFourierTransform[Psik[n, k], k, a, fp]], InverseFourierTransform[Psik[n, k], k, a, fp]]]];

(*--- D Psi/Da  ---*)
DPsi[n_, a_] := DPsi[n, a] = Simplify[D[Psi[n, a], a]];

(*--- Truncated sum of psi at order N  ---*)
SumPsi[N_, a_, \[Tau]_] := SumPsi[N, a, \[Tau]] = Sum[Psi[j, a]*\[Tau]^j, {j, 1, N}];

(*--- F_q(a) ---*)
F[n_, N_, a_, k_] := F[n, N, a, k] = Simplify[ SeriesCoefficient[Exp[-I*k*SumPsi[N, a, \[Tau]]], {\[Tau], 0, n}]];

(*--- Second term | Dt^{a,p} (p,q,N,k)  ---*)
dpat[p_, q_, N_, k_] := dpat[p, q, N, k] = p*FourierTransform[Psi[p, a]*(DPsi[q, a] + F[q, N, a, l]), a, k, fp] /. {l -> k};

(*--- Second term | D^{2,n} (k)  ---*)
d2n[n_, k_] := d2n[n, k] = Simplify[ ExpandAll[ Sum[If[i != (n - i), dpat[i, n - i, n - 1, k] + dpat[n - i, i, n - 1, k], dpat[i, i, n - 1, k]], {i, 1, f[n + 1], 1}]]];

(*--- Third term | D^{b,p} (a)  ---*)
dpb[p_, a_] := dpb[p, a] = Sum[If[i != (p - i), i*Psi[i, a]*DPsi[p - i, a] + (p - i)*Psi[p - i, a]*DPsi[i, a], i*Psi[i, a]*DPsi[i, a]], {i, 1, f[p + 1], 1}];

(*--- Third term | T^{3,n} (k)  ---*)
t3n[n_, k_] := t3n[n, k] = FourierTransform[ Sum[If[p != (n - p), (dpb[p, a] *F[n - p, n - 1, a, l] + dpb[n - p, a] *F[p, n - 1, a, l]), dpb[p, a] *F[p, n - 1, a, l]], {p, 1, f[n + 1], 1}], a, k, fp] /. {l -> k};

(*--- RHS of equation 19 ---*)
Tk[n_, k_] := Tk[n, k] = Simplify[ExpandAll[Vk[n - 1, k]/2 - d2n[n, k] - t3n[n, k]]];

(*------- COMPUTATIONS -------*)

(*----- Compute Psi_n(a) and Z_n(a,tau) up till order nmax and export to file -----*)
nmin=2;nmax=36; Timing[Monitor[Do[Psi[n,a], {n,nmin,nmax}], Row[{ProgressIndicator[n,{nmin,nmax}],n}," "]]];
zFull[a_,\[Tau]] = Collect[a + SumPsi[nmax, a, \[Tau]], \[Tau]];
Export["./Z_Full.txt", Simplify[zFull[a,\[Tau]]]];
    
(*----- Compute W_n(a,tau) up till order nmax and export to file -----*)
Wrecoftau[n_] := Normal[Simplify[Series[Integrate[Series[Sin[2 a]/D[zFull[a, t^2], a], {t, 0, n}], t], {t, 0, n}]/t]] /. t -> tau^(1/2);
Wres = Wrecoftau[2 nmax] /. {tau -> \[Tau]};
Export["~/W_Full.txt", Wres];

(*----- Compute V_n(a,tau) uptill order nmax and export to file -----*)
Vrecoftau[n_] := Normal[Simplify[ Series[D[zFull[a, \[Tau]], \[Tau]], {\[Tau], 0, n}]]];
Vres = Vrecoftau[nmax];
Export["./V_Full.txt", Vres];

\end{lstlisting}

\end{document}